\documentclass[12pt,a4paper,JHEP3]{article}

\pdfoutput=1
\usepackage{jheppub}
\usepackage{slashed,cancel,lscape,caption,array,graphicx,subcaption}
\usepackage[utf8]{inputenc}
\usepackage{amsmath}
\usepackage{nicefrac}  % for 1/2 
\usepackage{mathtools}
\usepackage{multirow}
\usepackage{booktabs} %nice tables 
\usepackage{chngcntr} 
\usepackage{nicefrac}
\usepackage{cleveref}
\usepackage[normalem]{ulem}
\usepackage{hhline}
\usepackage[table,dvipsnames]{xcolor}

\usepackage{tikz}
\usetikzlibrary{decorations.pathmorphing}
\usetikzlibrary{decorations.markings}
\usetikzlibrary{positioning, shapes, snakes, arrows}
\usetikzlibrary{patterns}

\tikzset{
    wl/.style={line width=1pt},
    graviton/.style={line width=.8pt, -latex,decorate, decoration={snake, segment length=4pt,amplitude=1.8pt, pre length=.15cm, post length=.25cm}},
    worldlineStatic/.style={dotted, line width=1pt},
	worldline/.style={gray, line width=1pt},
	worldlineBold/.style={black, line width=.6pt},
	zUndirected/.style={line width=1pt},
	zParticle/.style={line width=1pt,postaction={decorate},decoration={markings,mark=at position .6 with {\arrow[#1]{latex}}}},
	zParticleF/.style={line width=1pt,postaction={decorate}},
	cscalar/.style={line width=1pt,postaction={decorate},decoration={markings,mark=at position .6 with {\arrow[#1]{latex}}}},
	cscalar2/.style={line width=1pt,postaction={decorate},decoration={markings,mark=at position .8 with {\arrow[#1]{latex}}}},
	photon/.style={line width =.8pt, decorate, decoration={snake, segment length=4pt, amplitude=1.8pt,  pre length=.1cm, post length=.1cm}},
	photonRed/.style={red, line width =.8pt, decorate, decoration={snake, segment length=4pt, amplitude=1.8pt,  pre length=.1cm, post length=.1cm}},
	cross/.style={cross out, line width =.8pt, draw=black, minimum size=2*(#1-\pgflinewidth), inner sep=0pt, outer sep=0pt},
cross/.default={4pt}
}

\newcommand{\Z}{\mathcal{Z}}

\DeclareFontFamily{OT1}{pzc}{} 
\DeclareFontShape{OT1}{pzc}{m}{it}{<-> s * [1.350] pzcmi7t}{}
\DeclareMathAlphabet{\mathpzc}{OT1}{pzc}{m}{it}

\def\cL{\mathcal{L}}
\def\cN{\mathcal{N}}
\def\cO{\mathcal{O}}
\def\cD{\mathcal{D}}

\def\cM{\mathcal{M}}

\def\cZ{\mathcal{Z}}

\def\eps{\epsilon}

\def\d{\mathrm{d}}
\def\D{\mathrm{D}}

\def\E{\mathcal{E}}
\def\B{\mathcal{B}}

\def\mn{{\mu\nu}}
\def\ab{{\alpha\beta}}
\def\mnone{{\mu_1\nu_1}}

\def\mnn{{\mu_n\nu_n}}

\def\balpha{{\bar\alpha}}

\def\i\math

\def\bH{\hat{b}}

\def\dd{\delta\!\!\!{}^-\!}

\def\d{\mathrm{d}}
\def\eps{\epsilon}

\renewcommand{\i}{\ensuremath{\mathrm{i}}}
\renewcommand{\d}{\ensuremath{\mathrm{d}}}

\def\nn{\nonumber}

\def\eqn#1{eq.~\eqref{#1}}
\def\eqns#1#2{eqs.~\eqref{#1} and~\eqref{#2}}

\def\sec#1{section~{\ref{#1}}}

\def\rcites#1{refs.~\cite{#1}}

\newcommand{\be}{\begin{equation}}
\newcommand{\ee}{\end{equation}}
\newcommand{\ba}{\begin{align}}
\newcommand{\ea}{\end{align}}
\ifx\genfrac\sdflkaj\else\fi
\newcommand{\sfrac}[2]{{\textstyle\frac{#1}{#2}}}

\newcommand{\ga}{{\mathfrak a}}
\newcommand{\gz}{{\mathfrak z}}

\newcommand{\Hnm}{H_{\text{nm}}}

\def\centerarc[#1](#2)(#3:#4:#5){ \draw[#1] ($(#2)+({#5*cos(#3)},{#5*sin(#3)})$) arc (#3:#4:#5); }
\begin{document}

\begin{flushright}
\begingroup\footnotesize\ttfamily
	HU-EP-24/34-RTG\\
    QMUL-PH-24-28
\endgroup
\end{flushright}

\vspace{12mm}

\begin{center}
{\LARGE\bfseries 
Spinning bodies in general relativity from bosonic worldline oscillators
	%Bosonic worldline oscillators for \\ higher-spin massive bodies
    %Bosonic oscillators as worldline \\ spin degrees of freedom
\par}

\vspace{15mm}

\begingroup\scshape\large 
	Kays Haddad,${}^{1}$ 
	Gustav Uhre Jakobsen,${}^{1,2}$  Gustav~Mogull,${}^{1,2,3}$
	and Jan~Plefka${}^{1}$
\endgroup
\vspace{3mm}
					
\textit{${}^{1}$Institut f\"ur Physik und IRIS Adlershof, Humboldt-Universit\"at zu Berlin, \\ 10099 Berlin, Germany} \\[0.25cm]
\textit{${}^{2}$Max-Planck-Institut f\"ur Gravitationsphysik
(Albert-Einstein-Institut), \\14476 Potsdam, Germany } \\[0.25cm]
\textit{${}^{3}$Centre for Theoretical Physics, Department of Physics and Astronomy, \\ \!\!\!\!Queen Mary University of London, Mile End Road, London E1~4NS, United Kingdom}

\bigskip
  
\texttt{\small\{kays.haddad, gustav.uhre.jakobsen, jan.plefka@physik.hu-berlin.de, g.mogull@qmul.ac.uk\}}

\vspace{5mm}

\textbf{Abstract}\vspace{5mm}\par
\begin{minipage}{14.7cm}
Worldline quantum field theory (WQFT) has proven itself a powerful tool for classical two-body scattering calculations in general relativity.
In this paper we develop a new worldline action involving bosonic oscillators, which enables the use of the WQFT formalism to describe massive compact bodies to all orders in their spins.
Inspired by bosonic string theory in the tensionless limit, we augment traditional trajectory variables with bosonic oscillators 
capturing the spin dependence. We show its equivalence to the covariant phase space description of a spinning body in curved space and 
clarify the role of the spin-supplementary condition in a Hamiltonian treatment. Higher-spin Hamiltonians are classified to linear 
and quadratic order in curvature.
Finally, perturbative computations at 1PM order for arbitrary powers and orientations of spin and at 2PM up to quartic spin order are performed, recovering results from the literature.
\end{minipage}\par

\end{center}
\setcounter{page}{0}
\thispagestyle{empty}
\newpage
 
\tableofcontents

\section{Introduction}

The analysis of gravitational wave (GW) signals from compact-object binaries, such as spinning black holes and neutron stars, demands high-precision predictions of their gravitational two-body dynamics and emitted far-field waveforms. These predictions are crucial for interpreting data from present 
\cite{LIGOScientific:2016aoc,LIGOScientific:2017vwq,LIGOScientific:2021djp} and future \cite{LISA:2017pwj,Punturo:2010zz,Ballmer:2022uxx} gravitational wave observatories, enabling us to infer source parameters like masses and spins as well
as further intrinsic parameters discerning neutron stars from black holes. In the coming decade  the planned third-generation detectors will drastically increase observational accuracies, such that the need for theoretical predictions of the spinning gravitational two-body problem going beyond the present precision has become increasingly urgent~\cite{Borhanian:2022czq}.
In order to meet this theoretical challenge, a spectrum of perturbative approximation schemes have been developed for solving the gravitational two-body
problem: the post-Newtonian~\cite{Blanchet:2013haa,Porto:2016pyg,Levi:2018nxp} (assuming weak fields and low velocities) and the post-Minkowskian \cite{Kosower:2022yvp,Bjerrum-Bohr:2022blt,Buonanno:2022pgc,DiVecchia:2023frv,Jakobsen:2023oow} (assuming only weak fields) expansions on the purely analytical side, and the semi-numerical gravitational self-force \cite{Mino:1996nk,Poisson:2011nh,Barack:2018yvs,Gralla:2021qaf} (assuming a small mass ratio between the two objects) expansion. 
High-order results from these approaches, complemented by numerical relativity~\cite{Pretorius:2005gq,Boyle:2019kee,Damour:2014afa} and resummation techniques such as the effective-one-body formalism \cite{Buonanno:1998gg,Buonanno:2000ef}, 
are assembled into waveform models that form the foundation of GW data analysis
\cite{Buonanno:2007pf,Damour:2007yf,Nagar:2018zoe,Ramos-Buades:2023ehm,Pompili:2023tna,Buonanno:2024vkx,Buonanno:2024byg}.

Recently, rapid advancements have been made in the post-Minkowskian (PM) expansion, particularly relevant for the unbound or scattering  problem,
which is also the focus of this work.
These advancements have been driven largely by the application of modern quantum field theory techniques and Feynman integration methods:
integration by parts \cite{Laporta:2000dsw,Smirnov:2008iw,Gehrmann:1999as,Klappert:2020nbg,Lange:2021edb},
differential equations \cite{Gehrmann:1999as,Henn:2013pwa,Meyer:2017joq,Dlapa:2020cwj,Peraro:2019svx}, and the method of regions \cite{Beneke:1997zp}. 
These innovations have rapidly pushed the state-of-the-art from 3PM to 5PM,
employing both underlying worldline 
\cite{Kalin:2020mvi,Kalin:2020fhe,Kalin:2020lmz,Mogull:2020sak,Jakobsen:2021smu,Dlapa:2021npj,Dlapa:2021vgp,Mougiakakos:2021ckm,Riva:2021vnj,Dlapa:2022lmu,Dlapa:2023hsl,Liu:2021zxr,Mougiakakos:2022sic,Riva:2022fru,Jakobsen:2021lvp,Jakobsen:2021zvh,Jakobsen:2022fcj,Jakobsen:2022zsx,Jakobsen:2022psy,Shi:2021qsb,Bastianelli:2021nbs,Comberiati:2022cpm,Wang:2022ntx,Ben-Shahar:2023djm,Bhattacharyya:2024aeq,Jakobsen:2023ndj,Jakobsen:2023hig,Jakobsen:2023pvx}
and scattering amplitude 
techniques \cite{Neill:2013wsa,Luna:2017dtq,Kosower:2018adc,Cristofoli:2021vyo,Bjerrum-Bohr:2013bxa,Bjerrum-Bohr:2018xdl,Bern:2019nnu,Bern:2019crd,Bjerrum-Bohr:2021wwt,Cheung:2020gyp,Bjerrum-Bohr:2021din,DiVecchia:2020ymx,DiVecchia:2021bdo,DiVecchia:2021ndb,DiVecchia:2022piu,Heissenberg:2022tsn,Damour:2020tta,Herrmann:2021tct,Damgaard:2019lfh,Damgaard:2021ipf,Damgaard:2023vnx,Aoude:2020onz,AccettulliHuber:2020dal,Brandhuber:2021eyq,Bern:2021dqo,Bern:2021yeh,Bern:2022kto,Bern:2023ity,Damgaard:2023ttc,Brandhuber:2023hhy,Brandhuber:2023hhl,DeAngelis:2023lvf,Herderschee:2023fxh,Caron-Huot:2023vxl,FebresCordero:2022jts,Bohnenblust:2023qmy,Bern:2024adl,Alaverdian:2024spu,Akpinar:2024meg,Bohnenblust:2024hkw}.
The Worldline Quantum Field Theory (WQFT) method~\cite{Mogull:2020sak,Jakobsen:2021zvh,Jakobsen:2022psy},
which emphasises the calculation of classical scattering observables such as the momentum impulse and waveform,
has been particularly successful:
the conservative 5PM impulse up to first self-force (1SF) order --- a four-loop calculation ---
was recently obtained~\cite{Driesse:2024xad},
and the dissipative result is upcoming \cite{5pmrad}. 
Using scattering amplitudes the same 5PM-1SF order was recently attained in $\mathcal{N}=8$ supergravity \cite{Bern:2024adl}.

Spin is an essential property of all compact astrophysical objects, and a key observational target for GW detectors~\cite{Zackay:2019tzo,Huang:2020ysn}. In two-body 
systems with generic spin orientations, the orbital plane and object spins precess about the total angular momentum, leading to modulations in the GW signal. 
Improving the accuracy of waveform models for generic spin orientations is paramount for future GW searches, binary black hole formation studies, and tests of general relativity. 
As long as the two compact objects are well-separated, they may be efficiently modelled as spinning point particles moving along worldline trajectories that interact through the gravitational force. Traditionally, the effective worldline description captures the object's spin through a spin tensor $S^{\mu\nu}(\tau)$  and a co-rotating frame $\Lambda_{\mu}^{A}(\tau)$~\cite{Porto:2005ac,Levi:2015msa,Liu:2021zxr,Vines:2016unv}, augmenting the trajectory $x^{\mu}(\tau)$ and momentum $p_{\mu}(\tau)$ canonical pair.
However, as one is generally more interested in the spin tensor,
carrying the frame in an action-based formalism can be cumbersome ---
though not impossible, as demonstrated in ref.~\cite{Ben-Shahar:2023djm}.

The spin tensor $S^{\mu\nu}$ couples to the space-time curvature quantified by an infinite series of higher-dimensional worldline operators with free Wilson coefficients in the logic of effective field theory~\cite{Goldberger:2004jt,Porto:2016pyg,Levi:2018nxp}.
At linear order in curvature, these operators are well understood, with a single operator for each spin multipole, and with Wilson coefficients entering from the spin quadrupole order; contributions from the spin monopole and dipole are fixed by diffeomorphism invariance of the action \cite{Levi:2015msa}.
These linear-curvature Wilson coefficients attain known special values when the compact object in question is a Kerr black hole~\cite{Vines:2017hyw}.
Corrections involving coupling the spin degrees of freedom to higher powers of the curvature are being actively explored with increasing zeal \cite{Levi:2020uwu,Levi:2020lfn,Gupta:2020lnv,Aoude:2020ygw,Kim:2021rfj,Bern:2022kto,Kim:2022bwv,Mandal:2022ufb,Levi:2022rrq,Saketh:2022xjb}, including an enumeration and analyses of these corrections in the language of on-shell scattering amplitudes \cite{Aoude:2022trd,Bautista:2022wjf,Haddad:2023ylx,Cangemi:2023ysz,Cangemi:2023bpe}.
The values of Wilson coefficients at quadratic order in curvature which specialize to the Kerr-black-hole context is a vibrant and expanding line of research, though as yet an action describing all spin multipoles of a Kerr black hole to quadratic order in curvature is not known \cite{Ivanov:2022qqt,Bautista:2022wjf,Saketh:2023bul,Scheopner:2023rzp,Bautista:2023sdf}.

Three of the present authors \cite{Mogull:2020sak,Jakobsen:2021zvh,Jakobsen:2022psy},
together with Sauer and Steinhoff, have pioneered a particularly efficient approach
for representing the spin degrees of freedom of a compact binary encounter using a superparticle formalism.  It involves a worldline action which enjoys an extended $\mathcal{N}=2$ supersymmetry, where the 
worldline trajectory $x^{\mu}(\tau)$ is augmented by an anti-commuting complex worldline  vector $\psi^{\mu}(\tau)$ carrying the spin degrees of 
freedom \cite{Jakobsen:2021lvp,Jakobsen:2021zvh}: here the spin tensor is a composite object, $S^{\mu\nu}=-2\i m\bar\psi^{[\mu}\psi^{\nu]}$.
This approach is particularly amenable to the WQFT method for determining
momentum and spin change observables, as well as the waveform, at high orders in perturbation theory.
Non-spinning results for the deflection observables \cite{Bern:2021dqo,Bern:2021yeh,Dlapa:2022lmu,Dlapa:2023hsl,Damgaard:2023ttc} were extended to the spinning situation using this method at two-loop
\cite{Jakobsen:2022fcj,Jakobsen:2022zsx} and three-loop orders \cite{Jakobsen:2023ndj,Jakobsen:2023hig}.
Given also the recent work on 5PM~\cite{Driesse:2024xad,5pmrad},
this implies that in a physical PM counting~\cite{Rettegno:2023ghr,Buonanno:2024vkx},
where each order of spin contributes a PM order,
the complete 5PM dynamics for the deflection observables is now known up to 1SF order.

An important aspect of the description of the angular momentum of an extended body --- or spin ---
in general relativity is the so-called spin-supplementary condition (SSC).
In the multipolar description of extended bodies (see e.g. refs.~\cite{Dixon:1964cjb,DixonMultipoles:1967}), the SSC serves to close the otherwise-indeterminate equations of motion for the linear and angular momenta by imposing a further constraint on the angular momentum (spin) tensor $S^{\mu\nu}$.
Effectively, the six degrees of freedom of the spin tensor are reduced to three, which, in some reference frame, are encoded by the Pauli-Lubanski spin vector $\mathbf{S}$.
The SSC can be regarded as the fixing of a gauge freedom, any choice of which produces physically equivalent descriptions of the dynamics (though certain choices yield simplifications) \cite{Costa:2011zn,Costa:2014nta}.
In the superparticle formalism the supersymmetry itself provides the SSC constraint.
Yet, while elegant, this supersymmetric approach to spin remains
limited to spin-orbit and spin-spin interactions of the binary only,
as it remains unclear how to maintain the symmetries of the worldline action beyond $\cN=2$ supersymmetries in a gravitational background.
This has, so far, presented an obstacle to further scattering calculations using the WQFT.

In this paper, we therefore introduce a novel formalism that overcomes the limitations of the supersymmetric worldline.
That is, our new formulation is capable of describing spin interactions to \emph{any} order.
The key inspiration to do so comes from bosonic string theory in the tensionless  ($\alpha'\to\infty$)  limit: we introduce a bosonic spinning worldline model by augmenting the trajectory phase space variables $\{x^{\mu},p_{\nu}\}=\delta^{\mu}_{\nu}$ with a set of bosonic oscillators 
$\{\alpha^{a},\bar\alpha^{b}\}=-\i \eta^{ab}/m$  living in the flat tangent space. 
In great similarity to the supersymmetric model, we identify $S^{\mu\nu}=-2\i m\bar\alpha^{[\mu}\alpha^{\nu]}$. We derive this bosonic oscillator model from the covariant phase space description of a spinning particle in curved space-time introduced by d'Ambrosi, Kumar,  van de Vis and van Holten  in refs.~\cite{dAmbrosi:2015ndl,vanHolten:2015vfa,dAmbrosi:2015wqz}, which has been a great inspiration for our work.

The outline for our paper is as follows.
First, we construct a higher-spin Hamiltonian that preserves the common and convenient \emph{covariant} SSC.
Our procedure allows for the building of such a Hamiltonian to all spin and curvature orders, but we will focus here on all spin orders at linear order in curvature, and the leading physical PM order (up to quartic spins) at quadratic order in curvature.
By introducing the bosonic oscillators $\alpha^{\mu}$ and $\bar{\alpha}^{\mu}$,
a canonically-conjugate pair encoding the spin degrees of freedom,
we then Legendre transform to the Lagrangian, yielding an action suitable for WQFT perturbative calculations.
We then set up the perturbative framework of WQFT to perform higher-spin precision computations.
This affords us an opportunity to make contact with recent similar analyses which have explored the relaxation of the SSC constraint \cite{Bern:2020buy,Bern:2022kto,Bern:2023ity,Alaverdian:2024spu}.
As a proof of principle of 
our new approach, we use WQFT to reproduce the actions emerging from our Hamiltonian analysis, along the way computing the all-order-in-spin tree-level deflection observables and finding agreement with the literature.
At one-loop level we also include the first layer of quadratic-in-curvature spin couplings on the worldline
in our computations and demonstrate the covariant SSC preservation including these corrections explicitly,  reassuring us of its suitability for pushing the boundaries on predictions regarding the scattering of spinning compact bodies in classical general relativity.

\section{The Hamiltonian approach to relativistic spinning bodies}\label{sec:Hamiltonian}

In this section we review the Hamiltonian dynamics of massive spinning bodies in curved 
space-time
leading to the efficient covariant formulation of d'Ambrosi et.~al.~\cite{dAmbrosi:2015ndl,vanHolten:2015vfa,dAmbrosi:2015wqz}.
After discussing the covariant phase space, involving Poisson brackets,
we will use this to develop Hamiltonians that preserve the covariant SSC on the spin
tensor, thus keeping only the degrees of freedom necessary to describe a relativistic spinning body.

\subsection{Covariant phase space}

The phase space of a massive spinning body is described by:
its position $x^{\mu}$, canonical momentum $p_{\mu}$,
and antisymmetric spin tensor $S^{ab}$.
Here we distinguish between curved-space indices $\mu,\nu,\ldots$
and locally flat indices $a,b,\ldots$ taken in a frame defined by the vierbein $e_\mu^a(x)$.
In order to set up a Hamiltonian dynamical system,
we postulate the canonical set of Poisson brackets:
\begin{subequations}\label{eq:basicBrackets}
\begin{align}
\{ x^{\mu}, p_{\nu}\} &= \delta^{\mu}_{\nu}\, , \label{funpxpc} \\
\{ S^{ab} , S^{cd}\} & = 
\eta^{bd}\, S^{ac}+ \eta^{ac}\, S^{bd}
-\eta^{bc}\, S^{ad} -\eta^{ad}\, S^{bc}\,,\label{eq:spinBracket}
\end{align}
\end{subequations}
all other brackets vanishing.
The latter identifies $S^{ab}$ as enjoying the flat space-time SO(1,3) Lorentz algebra,
which includes both rotations and boosts.
As $\{(S^{ab})^2,\,\bullet\,\}=0$ in all cases, this also identifies
$s^2:=\frac{1}{2}S^{ab}S_{ab}$ as the usual Casimir invariant ---
later this will imply conservation of the length of the spin vector.
We now introduce the covariant momentum and spin tensor:
\begin{subequations}
\label{PiSdef}
\begin{align}
\pi_{\mu}&:= p_{\mu} + \sfrac{1}{2}\omega_{\mu,ab}S^{ab}\,,\label{eq:pidef}\\
S^{\mu\nu}&:=e^\mu_a e^\nu_bS^{ab}\, ,
\end{align}
\end{subequations}
where $\omega^{\mu}{}_{ab}$ denotes the spin-connection.
Using the brackets~\eqref{eq:basicBrackets},
it is then straightforward to derive the Poisson brackets for the set of covariant phase
space variables $\{x^{\mu},\pi_{\mu},S^{\mu\nu}\}$:
\begin{subequations}
\label{vHposbracks}
\begin{align}
\{ x^{\mu}, \pi_{\nu}\} & = \delta^{\mu}_{\nu}\, , \\
\{ S^{\mu\nu} , S^{\rho\kappa}\} & = 
g^{\nu\kappa}\, S^{\mu\rho}+ g^{\mu\rho}\, S^{\nu\kappa}
-g^{\nu\rho}\, S^{\mu\kappa} - g^{\mu\kappa}\, S^{\nu\rho}\, , \\
\{S^{\mu\nu}, \pi_{\rho}\} & = 2\Gamma^{[\mu}{}_{\rho\kappa}\, S^{\nu]\kappa}\, ,  \\
\{\pi_{\mu},\pi_{\nu}\} &= -\sfrac{1}{2} R_{\mu\nu ab}\, S^{ab}\, .\label{Spi}
\end{align}
\end{subequations}
Up to a convention-dependent sign in \eqn{Spi},
these are precisely the curved-space Poisson brackets of Ambrosi 
et.~al.~\cite{dAmbrosi:2015ndl,vanHolten:2015vfa,dAmbrosi:2015wqz} that were introduced as a consistent covariant canonical formalism for
spinning bodies allowing for different choices of the Hamiltonian.

The 3+3 components of the spin tensor $S^{ij}$ and $S^{0i}$ generate rotations and boosts respectively.
As we only seek to describe the three degrees of freedom of a spin vector $S^i=\sfrac12\eps^{ijk}S^{jk}$,
i.e.~compact rotating bodies,
we therefore set the latter three components $S^{0i}=0$ ---
a choice of SSC.
To covariantize this frame-dependent statement in four dimensions,
we introduce an inertial frame defined by a timelike vector $t^\mu=(1,\mathbf{0})$,
so that $S_{\mu\nu}t^\nu=0$.
The most convenient choice for $t^\mu$ for us will be the (unit-normalized) covariant momentum $\hat{\pi}^\mu:=g^{\mu\nu}\pi_\nu/|\pi|$, with $|\pi|=\sqrt{g^{\mu\nu}\pi_{\mu}\pi_{\nu}}$.
This choice of frame $t^\mu=\hat{\pi}^\mu$ defines the so-called covariant SSC.
To implement this, we decompose the spin tensor as \cite{dAmbrosi:2015ndl,vanHolten:2015vfa,dAmbrosi:2015wqz}
\begin{align}\label{eq:spinDecomp}
S^{\mu\nu} &= \varepsilon^{\mu\nu \rho\sigma}\hat{\pi}_{\rho}S_{\sigma} + Z^{\mu}\hat{\pi}^{\nu} -  Z^{\nu}\hat{\pi}^{\mu}\,.
\end{align}
where $\varepsilon_{\mu\nu\rho\sigma}:=\sqrt{-g}\eps_{\mu\nu\rho\sigma}$ is the covariant Levi-Civita tensor ($\eps_{0123}=1$),
whose indices may be raised and lowered using the metric $g_{\mu\nu}$.
Here we have introduced the spin vector $S^{\mu}$ and the SSC vector $Z^\mu$:
\begin{align}\label{aZdefs}
	S_{\mu}&:= \sfrac{1}{2}\varepsilon_{\mu\nu\rho\kappa}\hat{\pi}^{\nu}S^{\rho\kappa}\, , 
	&
	Z^{\mu}&:= S^{\mu\nu}\hat{\pi}_{\nu}\, \, .
\end{align}
Both vectors are orthogonal to $\pi^{\mu}$ by construction: $\pi\cdot S=\pi\cdot Z=0$.
Upon inverting these relationships one recovers the decomposition of the spin tensor~\eqref{eq:spinDecomp}.
The covariant SSC is then given by $Z^\mu=0$,
with the spin vector $S^\mu$ characterizing the remaining three spin degrees of freedom.

We note that other choices of the SSC are possible, though they will not be our main focus in this paper.
By introducing a body-fixed frame $\Lambda_{A}{}^{\mu}$,
one may for example consider the canonical (Pryce-Newton-Wigner) SSC~\cite{Pryce:1935ibt,Pryce:1948pf,Newton:1949cq} $S_{\mu\nu}(\hat{\pi}^\nu+{\Lambda_0}^\nu)=0$,
which is helpful for formulating a canonical phase-space algebra in the context of conservative (center-of-mass) two-body systems.
This requires an extension of the Poisson brackets~\eqref{vHposbracks} 
to include $\Lambda_{A}{}^{\mu}$ in the phase-space,
which we will briefly describe in \sec{sec:TransitionToLagrangian}.
One might also consider disgregarding the SSC altogether, as recently discussed in \rcites{Bern:2023ity,Alaverdian:2024spu},
and allowing the spin tensor $S^{\mu\nu}$ to carry all six degrees of freedom.
In this case, using
\begin{align}
	2s^{2}:=S^{\mu\nu}S_{\mu\nu} &= -2\left(S^2-Z^2\right)\,,
\end{align}
we see that conservation of $S^{\mu\nu}S_{\mu\nu}$ no longer implies conservation
of the length of the spin vector $S^\mu$.
The vector $Z^\mu$ may in this case be interpreted as a mass dipole moment,
describing the position of the body's center of mass relative to the worldline~\cite{Costa:2011zn,Costa:2014nta}.

\subsection{Preserving the spin-supplementary condition }\label{sec:HSHamiltonian}

Let us now define a Hamiltonian generating the time evolution along the
worldline of the massive spinning body.
Our intention is to preserve the SSC condition $Z^{\mu}=0$ on the spin tensor throughout this evolution.
The most general form of the total Hamiltonian of a spinning massive body takes the form
\begin{align}\label{eq:HTotal}
	H_T=eH[e_{\mu}^a(x),p_{\mu},S^{\mu}] +\zeta_\mu Z^\mu+a(S^{\mu\nu}S_{\mu\nu}-2s^2)\,,
\end{align}
where $H$ depends on spin only via the spin vector $S^\mu$, not the SSC vector $Z^\mu$.
We have also introduced the einbein $e$, the U(1) gauge field $a$ and the Lagrange multiplier $\zeta_\mu$,
whose roles are to enforce conservation of energy, spin length and SSC respectively.
As the Casimir invariant $S^{\mu\nu}S_{\mu\nu}$ has vanishing Poisson brackets,
we generally ignore the last term and gauge fix $a=0$ without loss of generality.

In order for the covariant SSC choice $Z^\mu=0$ to be preserved,
the essential requirement on the total Hamiltonian $H_T$ is that\footnote{
	Under the weak equality used here, there is no distinction between the $\dot{Z}^\mu=\d Z^{\mu}/\d\tau$ and the covariant
	time derivative $\D Z^\mu/\d\tau:=\dot{x}^\nu\nabla_\nu Z^\mu$.
}
\begin{align}
    \dot{Z}^\mu=\{Z^\mu,H_T\}\approx0\,.
\end{align}
Following Dirac's notation~\cite{dirac2001lectures},
the symbol $\approx$ here denotes a weak equality,
which is held to be true once we impose the vanishing of all conserved charges
$0=H=Z^\mu=(S^{\mu\nu}S_{\mu\nu}-2s^2)$
on the right-hand side ---
no such condition is imposed on a strong equality.\footnote{Said otherwise, weak equality 
implies that all initially vanishing charges stay zero under time evolution.}
This requirement provides us with sufficient constraints in order to solve for the necessary components of $\zeta_\mu$.
By doing so, we identify the SSC as a second-class constraint in the sense of Dirac~\cite{dirac2001lectures};
the energy and spin length conservation requirements,
which cannot be solved for in this way,
are first-class constraints.
We obtain:
\begin{align}\label{eq:chiConstraint}
	\{Z^\mu,H_T\}\approx e\{Z^\mu,H\}+\zeta_\nu\{Z^\mu,Z^\nu\}\,.
\end{align}
Using the Poisson brackets~\eqref{vHposbracks}, we have
\begin{align}
	\{Z^\mu,Z^\nu\}\approx S^{\mu\nu}-\sfrac1{2|\pi|^2}R_{\rho\sigma\lambda\tau}S^{\mu\rho}S^{\nu\sigma}S^{\lambda\tau}\, .
\end{align}
Introducing $\psi^\mu:=S^{\mu\nu}\zeta_\nu$
our essential requirement is therefore
\begin{align}\label{eq:psiSol}
	\psi^\mu+\sfrac1{2|\pi|^2}R_{\rho\sigma\lambda\tau}S^{\mu\rho}S^{\lambda\tau}\psi^\sigma\approx -e\{Z^\mu,H\}\,.
\end{align}
As we will generally solve for $\psi^\mu$ order-by-order in the spins, 
$\psi^{\mu}=\sum_{n>1}\psi^{\mu}_{(n)}$, we project this equation on a fixed spin order $n$:
\begin{align}\label{eq:psiSolorder}
	\psi_{(n)}^\mu+\sfrac 1{2|\pi|^{2}}R_{\rho\sigma\lambda\tau}S^{\mu\rho}S^{\lambda\tau}\psi_{{(n-2)}}^\sigma\approx -e\{Z^\mu,H\}|_{S^{n}}\,.
\end{align}
Then $\zeta_\mu Z^\mu=\zeta_\mu S^{\mu\nu}\hat{\pi}_\nu=-\psi^\mu\hat{\pi}_\mu$ may be inserted into $H_T$~\eqref{eq:HTotal}.
This provides a fully constructive mechanism for building up an SSC-preserving Hamiltonian,
regardless of what ansatz is made for $H$ at any order in spin or curvature.

One might object here, that solving \eqn{eq:psiSol} only weakly for $\psi^\mu$ is insufficient
as any linear-in-$Z$ terms would still be relevant at the level of the total Hamiltonian $H_T$.
However, one can straightforwardly show that this is not the case, as the subsequent contraction with $\hat\pi_{\mu}$
elevates these terms to quadratic order in $Z$. As a sanity check one may show that the constructed solution
for $\zeta\cdot Z$ inserted into $H_{T}$ indeed obeys $\{Z^{\mu}, H_{T}\}\approx 0$.

\subsection{Constituents of the Hamiltonian ansatz}

Having described our procedure for building up an SSC-preserving total Hamiltonian $H_T$~\eqref{eq:HTotal},
our next step will be to make a suitable ansatz for the $Z$-independent contribution $H$.
This ansatz includes the spin vector $S^\mu$ and momentum $\pi_\mu$ coupled to the curvature of the gravitational field.
In the following we will ignore couplings to the Ricci tensor and Ricci scalar, which amounts to
an assumption that the bodies live in a vacuum spacetime.\footnote{We have observed that such couplings do not make a difference at the formal 2PM order to be considered below.}
Thus, the gravitational dynamics of the worldline are encoded in the Weyl tensor $C_{\mu\nu\rho\kappa}$,
which may be decomposed in an analogous manner to the spin tensor~\eqref{eq:spinDecomp}.
First, we introduce the electric and magnetic curvature tensors:
\begin{subequations}\label{eq:EBdef}
\begin{align}
    E_\mn+\i
    B_\mn
   &=
   (C_{\mu\alpha\nu\beta}+\i
   C^*_{\mu\alpha\nu\beta}) \hat{\pi}^{\alpha} \hat{\pi}^{\beta}\\
   &=
   \frac18{G_{\mu\alpha}}^{\rho\sigma}{G_{\nu\beta}}^{\lambda\tau}C_{\rho\sigma\lambda\tau}\hat{\pi}^\alpha\hat{\pi}^\beta\,,\label{eq:EBextra}
\end{align}
\end{subequations}
which depend on the dual Weyl tensor $C^*_{\mn\ab}=\frac12 \varepsilon_\mn{}^{\rho\sigma} C_{\rho\sigma\ab}$ and
\begin{align}
    G_{\mn\rho\sigma}=2g_{\mu[\rho}g_{\sigma]\nu}+\i \varepsilon_{\mn\rho\sigma}\,.
\end{align}
By inverting the relationship in \eqn{eq:EBdef} we obtain a decomposition of the the Weyl tensor
into the instantaneous frame of the point particle (in four dimensions) \cite{Vines:2016unv}:
\begin{align}\label{eq:WeylInstFrameDecomp}
    C_{\mu\nu\rho\kappa} + 
    \i 
    C^{\ast}_{\mu\nu\rho\kappa} 
    =
    G_{\mu\nu}{}^{\alpha\sigma}G_{\rho\kappa}{}^{\beta\epsilon}
    \left
     (
         E_{\sigma\epsilon} 
        + \i 
         B_{\sigma\epsilon}
     \right ) \hat{\pi}_{\alpha}\hat{\pi}_{\beta}\,.
\end{align}
To derive this, one inserts the definition of the $E_{\mu\nu}$ and $B_{\mu\nu}$ tensors~\eqref{eq:EBextra}
into the right-hand side above, making use of another useful identity (in four dimensions):
\begin{align}
    G_{\mu\nu\rho\sigma}={G_{\mu\nu\lambda}}^\alpha{G_{\rho\sigma}}^{\lambda\beta}\hat{\pi}_\alpha\hat{\pi}_\beta\,.
\end{align}
This follows from the fact that $\varepsilon_{[\mu\nu\rho\sigma}\pi_{\lambda]}=0$.
 Finally, taking the real part of \eqn{eq:WeylInstFrameDecomp}, we obtain the following explicit relationship:
\begin{align} \label{poprel}
\begin{aligned}
 C_{\mu\nu\rho\kappa} =&
     - 2E_{\mu[\rho}(g_{\kappa]\nu}-2\hat\pi_{\kappa]} \hat\pi_{\nu})
     + 2E_{\nu[\rho}(g_{\kappa]\mu}-2\hat\pi_{\kappa]} \hat\pi_{\mu})
    \\ &
        + 2B_{[\mu  }{}^{\lambda} \hat\pi_{\nu]}\hat\pi^\alpha\varepsilon_{\alpha\rho \kappa  \lambda} 
        + 2B_{[\rho }{}^{ \lambda} \hat\pi_{\kappa]}\hat\pi^\alpha\varepsilon_{\alpha\mu \nu  \lambda} 
     \,,
\end{aligned}
\end{align}
expressing the Weyl tensor in terms of the gravito-electromagnetic tensors.

Constructing an ansatz for the Hamiltonian capturing spinning dynamics at all multipole orders entails
the enumeration of an infinite number of interaction terms.
Power counting aids us in this task by classifying operators in terms of the characteristic size of their contributions to dynamics.
Our dimensional conventions are $c=1$, equating
length and time, but keeping $\hbar$ dimensionful as we are doing classical physics.  Hence, we discriminate between mass and length scales. 
To this end, we assume that the only length scale is
the Schwarzschild radius $2G m$ --- a compact body assumption, valid for black holes and
(as an order of magnitude) for
neutron stars.
This scale must balance all lengths 
associated with the curvature tensors and their covariant derivatives to render the Wilson coefficients dimensionless.
The building blocks of the EFT are then the rescaled dimensionless curvature tensors
$(Gm)^2 E_\mn$ and $(Gm)^2 B_\mn$, rescaled covariant derivatives $(Gm) \nabla_\mu$, 
the covariant momentum $\pi_\mu$ and the dimensionless spin vector:
\begin{align}
    \chi^\mu=\frac{S^\mu}{Gm^2}\,.
\end{align}
The former curvature tensors and derivatives appear in the specific combinations
\begin{align}
    &(Gm)^{n+2}\nabla_{\mu_1}\ldots \nabla_{\mu_n} E_{\ab}
    \ ,&
    &(Gm)^{n+2}\nabla_{\mu_1}\ldots \nabla_{\mu_n} B_{\ab}
    \ ,
\end{align}
with the covariant derivatives acting only on the curvature tensor, not the momenta $\pi^\mu$. 
For example,
\begin{align}
    \nabla_\mu E_\ab = \nabla_\mu C_{\alpha\hat{\pi}\beta\hat{\pi}}=\hat{\pi}^\rho\hat{\pi}^\sigma \nabla_\mu C_{\alpha\rho\beta\sigma}\,,
\end{align}
using Schoonschip notation \cite{Strubbe:1974vj,Veltman:1991xb}, e.g.~$B_{SZ}=S^{\nu}Z^{\rho} B_{\nu\rho}$.
Power counting rules in physical PM, curvature and spin orders are given in table~\ref{table:scalings}.
Notice that, in the physical PM counting,
powers in the dimensionful spin vector $S^\mu$ are also included.
A separate power counting in terms of the spin is therefore often unnecessary,
and is itself only meaningful when $|\chi|\ll1$.
For black holes, $|\chi|\leq1$.

In the following subsections we will build up a Hamiltonian ansatz order-by-order in the curvature tensors.
In the broader context of a two-body system, which also includes a second body with mass $m_*\ll m$,
the $n$th order in curvature is sufficient to describe the system up to $(n-1)$ orders beyond
leading in the mass ratio $m_*/m$.
One can infer this straightforwardly from the Feynman diagrams involved:
at leading order in the mass ratio (static limit) one requires only single-graviton couplings,
at sub-leading order one requires two-graviton couplings, and so on.

\begin{table}
    \centering
\begin{tabular}{|m{6em}|m{5em}|m{5em}|m{5em}|}
  \hhline{~|-|-|-|}
  \multicolumn{1}{c|}{} &
  \cellcolor{lightgray}  \hfil $(Gm)^2 E_\mn $ &
  \cellcolor{lightgray} \hfil $(Gm) \nabla_\mu$ &
  \cellcolor{lightgray} \hfil $\chi^\mu$ 
  \\ 
  \hline
  \cellcolor{lightgray} \hfil Physical PM &
  \hfil $\sim\epsilon_{\rm PM}^2$ &
  \hfil $\sim\epsilon_{\rm PM}$ &
  \hfil $\sim 1$ 
  \\ 
  \hline
  \cellcolor{lightgray} \hfil Curvature &
  \hfil $\sim\epsilon_{\rm R}^{\hphantom{2}}$ &
  \hfil $\sim 1$ &
  \hfil $\sim 1$ 
  \\
  \hline
  \cellcolor{lightgray} \hfil Spin &
  \hfil $\sim 1$ &
  \hfil $\sim 1$ &
  \hfil $\sim \eps_{\chi}$ 
  \\
  \hline
\end{tabular}
\caption{
    Scalings of dimensionless variables in the physical PM, curvature and spin regimes.
    Here $\epsilon_{\ast}\ll1$ are dimensionless power-counting parameters. In the context of a two-body system the
    curvature expansion coincides with a gravitational self-force (GSF) expansion in the mass ratio.
}
\label{table:scalings}
\end{table}

\subsection{SSC-preserving linear-in-curvature Hamiltonian}\label{sec:SSClincurv}

To begin with, let us consider couplings up to linear order in curvatures, but arbitrarily high spin powers,
restricting ourselves to parity-even contributions.
Such a description is entirely sufficient to describe a static object.
The relevant EFT analysis was first carried out in ref.~\cite{Levi:2015msa},
and leads to a single infinity of effective couplings:
\begin{align}\label{eq:spinCorrections}
	H_{RS^n}=
    \frac{m^{2-n}}{n!}
	\begin{cases}
		(-1)^{\frac{n}2}C_{ES^n}
        \big(\nabla_S\big)^{n-2}
         E_{SS}\,,& n\text{ even}\,,\\
        (-1)^{\frac{n-1}2}C_{BS^n}
        \big(\nabla_S \big)^{n-2}
        B_{SS}\,,& n\text{ odd}\,.
	\end{cases}
\end{align}
For generality, the Wilson coefficients may depend on the spin scalar $\chi^2$ such that $C_{ES^n}=C_{ES^n}(\chi^2)$.
Indeed, the result of Ref.~\cite{Levi:2015msa} was that any other coupling at linear order in curvature may be reduced to the ones above assuming the Einstein vacuum equations.
It is common to include the higher spin operators through the dynamical mass $\cM^2=m^2-2\sum_{n>1}H_{RS^n}$.
The linear-in-$R$ Hamiltonian then reads:
\begin{align}\label{Msqrdef}
	H=\sfrac{1}{2}\left(g^{\mu\nu}\, \pi_{\mu}\, \pi_{\nu}-\cM^2\right)=
    \sfrac{1}{2}\left(g^{\mu\nu}\, \pi_{\mu}\, \pi_{\nu}-m^2\right)+\sum_{n>1}H_{RS^n}\,.
\end{align}
Note, that for Kerr black holes, the Wilson coefficients take the  value one \cite{Levi:2015msa,Vines:2017hyw}.

We now solve for the SSC order-by-order in spins,
and thus write down the total Hamiltonian $H_T$.
The leading-order solution to \eqn{eq:psiSolorder} ($n=2$) is
\begin{align}\label{psi2full}
    \psi_{(2)}^{\mu} &\approx -e\!\left.\{Z^{\mu},H\}\right|_{S^2} \approx
    -\sfrac{e}{2} R_{\hat \pi \nu\alpha\beta} S^{\mu\nu} S^{\alpha\beta}\,.
\end{align}
As we will demonstrate shortly,
none of the higher-spin corrections~\eqref{eq:spinCorrections} play a role
in the SSC term until cubic order in spins. 
Using the spin tensor decomposition~\eqref{eq:spinDecomp},
\begin{align}
    \begin{aligned}
    \label{psi2a}
        \zeta\cdot Z= -\psi_{(2)}^\mu\hat{\pi}_\mu+\cO(S^3)
        &= - \sfrac{e}{2}R_{\hat{\pi} Z \alpha\beta}S^{\alpha\beta} +\cO(S^3)\\
        &=e\left(E_{ZZ}-B_{SZ}\right)
        +\cO(S^3)\,.
\end{aligned}
\end{align}
To reproduce the second line, one decomposes all dependence on curvature tensors
into $E_{\mu\nu}$ and $B_{\mu\nu}$ using \eqn{poprel}, thereby
ignoring all terms that vanish on support of the vacuum Einstein equations $R_{\mu\nu}=0$;
all spin dependence is converted into $S^\mu$ and $Z^\mu$ via \eqn{eq:spinDecomp}.
Thus, we may ensure conservation of the SSC by including
$-eB_{SZ}$ in the total Hamiltonian $H_T$.
The $E_{ZZ}$ term may be ignored, as it always weakly vanishes after taking a Poisson bracket.

Proceeding beyond quadratic order in spin,
working up to linear curvatures \eqn{eq:psiSolorder} reduces to
\begin{align}
    \psi_{(n)}^\mu\approx-e\{Z^\mu,H\}|_{S^n}+\cO(R^2)\,.
\end{align}
One may show that
\begin{align}
	\{Z^\mu,H_{RS^n}\}\approx
    \frac{m^{2-n}}{n!}
	\begin{cases}
		-(-1)^{\frac{n}{2}}C_{ES^n}|\pi|^{-1}S^{\mu\nu}\nabla_\nu(\nabla_S)^{n-2}E_{SS}+\cO(R^2)\,,& n\text{ even},\\
		-(-1)^{\frac{n-1}{2}}C_{BS^n}|\pi|^{-1}S^{\mu\nu}\nabla_\nu(\nabla_S)^{n-2}B_{SS}+\cO(R^2)\,,& n\text{ odd}.
	\end{cases}
\end{align}
In order to derive this, we have made use of the fact that
$\pi_\mu\{S^{\mu\nu},S^\sigma\}\sim\cO(R)$.
As $H_T=eH+\zeta\cdot Z$,
we can therefore now write down an all-order-in-spin SSC-preserving Hamiltonian up to linear order in curvature:
\begin{subequations}\label{eq:AllSpinHamiltonian}
\begin{align}
    H_T&=\frac{e}2\Big(g^{\mu\nu}\pi_\mu\pi_\nu-m^2-2B_{SZ}+2\sum_{n>1}\tilde{H}_{RS^n}
    \Big)\,,\\
	\tilde{H}_{RS^n}&=
    \frac{m^{2-n}}{n!}
	\begin{cases}
		(-1)^{\frac{n}{2}}C_{ES^n}(1+|\pi|^{-1}\nabla_Z)(\nabla_S)^{n-2}E_{SS}\,,& n\text{ even},\\
		(-1)^{\frac{n-1}{2}}C_{BS^n}(1+|\pi|^{-1}\nabla_Z)(\nabla_S)^{n-2}B_{SS}\,,& n\text{ odd}.
	\end{cases}\label{eq:Htildelin}
\end{align}
\end{subequations}
The clear pattern is: for every term that we include at a given order in spin $s$,
we must include a corresponding term at one order \emph{higher} in the spins $s+1$,
in order to compensate for any SSC violations.
One can straightforwardly check, and we have confirmed,
that $\{Z^\mu,H_T\}\approx\cO(R^2)$.

\subsection{Induced SSC-preserving higher-in-curvature terms}\label{sec:SSCR2Ind}

We have now constructed an all-order-spin Hamiltonian which describes a spinning body obeying
the covariant SSC up to linear-in-curvature precision.
The full gravitational dynamics of such an object, however, also depends on higher-curvature corrections.
The topic of higher-curvature operators is a rich and timely one involving the encoding of finite-size effects in the scattering (see, e.g., refs.~\cite{Haddad:2020que,Aoude:2022trd,Bern:2022kto,Ivanov:2022qqt,Cangemi:2022bew,Bautista:2022wjf,Haddad:2023ylx,Saketh:2023bul,Jakobsen:2023pvx,Bautista:2023sdf,Cangemi:2023bpe}).
In this paper we will consider quadratic-in-curvature corrections up to fourth order in the spin tensor.
In this subsection, we derive the quadratic-in-curvature terms induced by SSC preservation from the linear-in-curvature
couplings~\eqref{eq:spinCorrections}, first appearing (as we shall demonstrate) at quartic order in spin.
In the next subsection, we will extend $\mathcal{M}^{2}$~\eqref{Msqrdef} to include
quadratic powers of $E_{\mu\nu}$ and $B_{\mu\nu}$, with a corresponding new set of Wilson coefficients.

Our task is to solve \eqn{eq:psiSolorder} up to quartic spins, and deduce the quadratic-in-curvature corrections.
The leading-order solution $(n=2)$ was already derived in \eqn{psi2full},
and remains valid at higher orders in curvature.
At cubic order we insert the leading-spin term $-\frac{1}{2} C_{ES^{2}} E_{SS}$ 
of $\sum_{n>1}H_{RS^{n}}$ on the right-hand side
of \eqn{eq:psiSolorder} for $H$. The relevant Poisson bracket relation is\footnote{To be precise, one -- as always -- first evaluates the Poisson bracket $\{Z^{\mu},E_{SS}\}$ weakly and thereafter contracts the result with $\hat\pi_{\mu}$. }
\begin{align}
 \label{PBZEsspi}
\hat\pi_{\mu}\,  \{ Z^{\mu}, E_{SS} \} &= |\pi|^{-1}\nabla_{Z} E_{SS} +2(E_{SS} B_{ZS} -  Z\cdot S\, E_{S\nu} B_{S}{}^{\nu})|\pi|^{-2} + \cO(Z^{2})\, . 
\end{align}
At cubic order in spin only the term $\nabla_{Z} E_{SS}$ contributes, and we deduce 
\be\label{chi3sol}
\zeta^{(3)}_{\mu}Z^{\mu} = -\hat\pi_{\mu}\psi_{(3)}^{\mu} =  
-e\frac{ C_{ES^{2}}}{2}  \hat\pi_{\mu}\,  \{ Z^{\mu}, E_{SS} \} \Bigr|_{S^{3}}=
- e\frac{C_{ES^{2}}}{2|\pi|}\nabla_{Z}E_{SS} +\cO(Z^{2})\,.
\ee
Again, there is no change here from the previous subsection.

At fourth order in spins, quadratic-in-curvature terms
emerge from two sources: the quartic spin terms in the bracket \eqref{PBZEsspi},
and the iteration of the second term on the left-hand side of \eqn{eq:psiSolorder} with $\psi^{\sigma}_{(2)}$ of~\eqref{psi2full} inserted.
The latter is
\be
\hat\pi_{\mu}\sfrac{1}{2|\pi|^{2}}R_{\rho\sigma\lambda\tau} S^{\mu\rho}S^{\lambda\tau}\psi_{{(2)}}^\sigma
=  e(E_{SS} B_{ZS} -  Z\cdot S\,  E_{S\nu} B_{S}{}^{\nu})|\pi|^{-2}\, .
\ee
It is nice to see that the same quadratic curvature terms appear here as in \eqn{PBZEsspi}.
Combining these results we thus learn that at fourth order in spin the SSC-preserving $\zeta\cdot Z$ term in $H_{T}$ becomes
\begin{align}\label{eq:InducedR2}
\begin{split}
\zeta^{(4)}_{\mu}Z^{\mu} =-\hat\pi_{\mu}\psi_{(4)}^{\mu} = & - e\frac{C_{BS^{3}}}{6|\pi| m}\nabla_{Z}\nabla_{S}B_{SS} \\ &
-e\frac{C_{ES^{2}} -1}{|\pi|^{2}}\, (E_{SS} B_{ZS} -  Z\cdot S\, E_{S\nu} B_{S}{}^{\nu})
+\cO(Z^{2})\, .
\end{split}
\end{align}
The first term arises from $\hat\pi_{\mu}\,  \{ Z^{\mu}, \frac{\nabla_{S}}{m}B_{SS} \}$ for the cubic spin contribution,
and was already included in \eqn{eq:AllSpinHamiltonian}.
The second line represents the leading curvature-squared terms needed to secure an SSC-preserving theory up to quartic order in spin.
Notice that as the Wilson coefficient appears in the combination $C_{ES^{2}} -1$,
this implies that for a Kerr BH, where $C_{ES^2}=1$, this term does not contribute.

\subsection{Generic higher-in-curvature terms}\label{sec:SSCR2Gen}

    Besides the induced higher-in-curvature terms from the SSC-preservation mechanism, we may also add a new
layer of curvature-squared terms to $H$. Again, the parametrization in terms of the minimal set of 
operators $\{E_{\mu\nu}, B_{\mu\nu}, S^{\mu}\}$ is extremely useful.
At quadratic order in curvature a new wealth of possible couplings appears.
In order to consider only a finite number of couplings we assume a physical PM expansion. 
We consider, then, all terms to quartic physical PM order:
\begin{subequations}\label{eq:R2S4Ops}
\begin{align}
    H_{R^{2}} 
    &=
    (Gm)^{4} m^{2}\Big(
           C_{E^2}
         E_{\mu\nu}E^{\mu\nu} 
       + 
       C_{B^2}
       B_{\mu\nu}B^{\mu\nu} \Big)
     \,,\\
    H_{R^{2}S^2} 
    &=
    (Gm)^{2}\Big( 
       C_{E^2S^2}
       E_{S\mu}E_{S}{}^{\mu}
       +
       C_{B^2S^2}
       B_{S\mu}B_{S}{}^{\mu}\Big)
     \,,\\
     H_{R^{2}S^4} 
    &=
    m^{-2}\Big( 
       C_{E^{2}S^{4}}
       E_{SS}E_{SS} 
       +
       C_{B^{2}S^{4}}
       B_{SS}B_{SS} 
       \big)
     \, .
\end{align}
\end{subequations}
Using $S^\mu=Gm^2\chi^\mu$ makes the physical PM counting manifest ---
all three terms carry $(Gm)^4$.
Again, the Wilson coefficients may depend on $\chi^2$
such that $C_{E^2 S^n}=C_{E^2 S^n}(\chi^2)$ and $C_{B^2S^n}=C_{B^2 S^n}(\chi^2)$.
Assuming a further expansion in $|\chi|\ll1$,
we may further expand these interactions up to fourth order in spin:
\begin{subequations}
\begin{align}
    C_{E^2}(\chi^2) 
    &= C_{E^2}^{(0)}
    +
    \chi^2 C_{E^2}^{(1)}
    +
    \chi^4 C_{E^2}^{(2)}
    +\dots\,,
    \\
    C_{E^2S^2}(\chi^2) 
    &= C_{E^2S^2}^{(0)}
    +
    \chi^2 C_{E^2S^2}^{(1)}
    +\dots\,,
    \\
    C_{E^2S^4}(\chi^2) 
    &= C_{E^2S^4}^{(0)}
    +\dots\,.
\end{align}
\end{subequations}
In a scattering computation, these couplings contribute to an observable,
such as the impulse or scattering angle, at the \emph{physical} 6PM order.
This is because each of the two emitted gravitons must
be absorbed by a second worldline, yielding two additional factors of $G$ in the computation of physical observables.
If we also restrict our attention to Kerr black holes, then all $C_{(E/B)^{2}S^{n}}^{(m)}$ with $m+n\leq4$ are vanishing. More specifically, the constraints on $C_{(E/B)^{2}S^{n}}^{(4-n)}$ depend on the linear-in-curvature basis used and were determined in our case by matching to known Kerr observables \cite{Bautista:2022wjf,Bautista:2023sdf}.
The vanishing of the rest of the coefficients for Kerr black holes has been demonstrated in ref.~\cite{Saketh:2023bul}.

Again, \cref{eq:R2S4Ops} will need to be augmented by terms linear
in $Z^{\mu}$ in order to preserve the SSC. In complete analogy to the analysis in section \ref{sec:SSClincurv},
this will --- at the quadratic order in curvature --- yield the contributions
\be\label{eq:R2GeneralInducedOps}
\sum_{n\geq 0}\tilde H_{R^{2}S^{n}} = \sum_{n\geq 0}(1+ |\pi|^{-1} \nabla_{Z}) H_{R^{2}S^{n}} +\cO(R^{3})\, .
\ee
We refrain here from working out the associated cubic-in-curvature terms.
Our complete quadratic-in-curvature Hamiltonian, including terms up to quartic powers in spin, therefore takes the form
\begin{align}
\begin{aligned}
    H_T&=\frac{e}2\Big[g^{\mu\nu}\pi_\mu\pi_\nu-m^2-2B_{SZ}
    -\frac{2(C_{ES^2}-1)}{|\pi|^2}(E_{SS} B_{ZS} -  Z\cdot S\, E_{S\nu} B_{S}{}^{\nu})\Big]\\
    &\qquad+e\sum_{n>1}(\tilde{H}_{RS^n}+\tilde{H}_{R^2S^n})\,.
\end{aligned}
\end{align}
Expressions for $\tilde{H}_{RS^n}$ and $\tilde{H}_{R^2S^n}$ were provided in \eqns{eq:Htildelin}{eq:R2GeneralInducedOps} respectively.

\section{The Lagrangian approach to relativistic spinning bodies}\label{sec:TransitionToLagrangian}

In order to set up an efficient
Feynman diagrammatic perturbation theory enabling us to compute observables,
we now intend to transition from the Hamiltonian to the Lagrangian second-order formalism.
In principle, this is a simple task: Legendre transform the full Hamiltonian of the system~\eqref{eq:HTotal}
to give a worldline Lagrangian,
and then solve for the canonical momentum $\pi_\mu$ in terms of $\dot{x}^\mu$.
The main practical difficulty is that we cannot write down a worldline action that involves only
$x^\mu$, $\pi_\mu$ and $S^{\mu\nu}$ (plus the metric and curvature tensors)
which are the essential degrees of freedom, the latter restricted by the SSC.
The issue is the kinetic term for the spin degrees of freedom, that cannot be written
down with $S^{\mu\nu}$ alone -- we must introduce additional 
variables into our system, or represent $S^{\mu\nu}$ as a composite object.
We briefly review two established approaches for doing so before moving to the central innovation of our work:
the representation of the spin tensor in terms of bosonic oscillators.

\subsection{The body-fixed frame approach}

The most conventional solution within worldline EFT is to use the body-fixed frame $\Lambda_{A}{}^{\mu}(\tau)$,
which satisfies $\eta^{AB}\Lambda_{A}{}^{\mu}\Lambda_{B}{}^{\nu}=g^{\mu\nu}$.
From this, one may define the antisymmetric angular-velocity tensor
\begin{align}
	\Omega^{\mu\nu}=\Lambda_{A}{}^{\mu}\frac{\D\Lambda^{A\nu}}{\d\tau}\,,
\end{align}
which is dual to the spin tensor: $\{S_{\mu\nu},\Omega^{\rho\sigma}\}=\delta^{[\rho}_\mu\delta^{\sigma]}_\nu$.
The first-order form of the action of the full system is then given by
\begin{align}
	S=-\int\!\d\tau\left(p_\mu\dot{x}^\mu+\frac12S_{ab}\Omega^{ab}-H_T\right)\,.
\end{align}
The main advantage of this approach is that it leaves flexibility over the choice of SSC:
by using the body-fixed frame, the components $S^{0i}$ may be taken to vanish in alternative frames of reference.
See e.g.~\cite{Vines:2016unv,Porto:2016pyg,Levi:2018nxp} for more details on this approach, and in
particular \cite{Ben-Shahar:2023djm} for a diagrammatic implementation using the WQFT formalism.
In this paper though, we lay out a more economical mechanism avoiding the need for a dynamical body fixed frame  for handling spin degrees of freedom,
motivated by our desire to specialize to the covariant SSC.

\subsection{The $\cN=2$ supersymmetric formalism}

More recently, three of the present authors with Steinhoff introduced an economical 
worldline formulation of spin using the
$\mathcal{N}=2$ superparticle~\cite{Jakobsen:2021lvp,Jakobsen:2021zvh},
building upon the works of \rcites{Gibbons:1993ap,Bastianelli:2005vk,Bastianelli:2005uy}.
Here the spin is carried by a Grassmann-odd
complex vector $\psi^{a}(\tau)$ on the worldline,
with the fundamental Poisson-bracket:
\begin{align}
	\{\psi^a,\bar{\psi}^b\}=-\i \eta^{ab}\,.
\end{align}
This ensures that the spin tensor $S^{ab}=-2\i\bar{\psi}^{[a}\psi^{b]}$,
considered as a composite object, enjoys the brackets~\eqref{eq:spinBracket}.
Within this approach, preservation of the covariant SSC is ensured by the existence
of an $\cN=2$ supersymmetry, with charges
\begin{subequations}
\begin{align}
	H&=\sfrac12(g^{\mu\nu}\pi_\mu\pi_\nu-m^2-R_{abcd}\bar{\psi}^a\psi^b\bar{\psi}^c\psi^d)\,,\label{eq:susyH}\\
	J&=\eta_{ab}\bar{\psi}^a\psi^b\,,\\
	Q&=\psi^a e_a^\mu\pi_\mu\,,\\
	\bar{Q}&=\bar{\psi}^a e_a^\mu\pi_\mu\,.
\end{align}
\end{subequations}
These obey the $\cN=2$ supersymmetry algebra (all other brackets vanishing):
\begin{align}\label{eq:susyAlgebra}
	\{Q,\bar{Q}\}&=-2iH\,, &
	\{J,Q\}&=-iQ\,, &
	\{J,\bar{Q}\}&=i\bar{Q}\,,
\end{align}
The existence of this algebra implies that $H$, $J$, $Q$ and $\bar{Q}$ form a set of first-class constraints,
as the brackets of all four supercharges weakly vanish.
Preservation of the covariant SSC is then guaranteed as 
$Z^\mu=S^{\mu\nu}\hat{\pi}_\nu=-i(\bar{\psi}^{\mu}Q+\psi^{\mu}\bar{Q})/|\pi|\approx0$.

In this $\cN=2$ SUSY formulation, there is no need to solve for the SSC explicitly as we did in \sec{sec:HSHamiltonian}  ---
the automatically SSC-preserving Hamiltonian~\eqref{eq:susyH} can instead be derived from the bracket $\{Q,\bar{Q}\}$.
Decomposing the third term of this Hamiltonian onto the $\{E_{\mu\nu}, B_{\mu\nu}, S^{\mu},Z^\mu\}$ basis
(in a vacuum space-time)
\begin{align}\label{RSSidentity}
    R_{abcd}\bar{\psi}^a\psi^b\bar{\psi}^c\psi^d=-\sfrac14R_{\mu\nu\rho\sigma}S^{\mu\nu}S^{\rho\sigma}
    =E_{SS}-E_{ZZ}+2B_{SZ}\, ,
\end{align}
shows that the $B_{SZ}$ term appearing here is exactly the one in the SSC-preserving Hamiltonian~\eqref{eq:AllSpinHamiltonian}.
Meanwhile, the extra $E_{SS}$ term induces a shift $C_{ES^2}\to\tilde{C}_{ES^2}=C_{ES^2}-1$
compared with $H_{RS^2}$~\eqref{eq:spinCorrections} --- and the $E_{ZZ}$ term is irrelevant.
Thus, the unique $\cN=2$ SUSY theory wherein $\tilde{C}_{ES^2}=0$ represents black holes,
whereas $\tilde{C}_{ES^2}\neq0$ represents more generic compact bodies.
In this case, the $\cN=2$ SUSY is obeyed only ``approximately'',
i.e.~up to and including quadratic order in spin.

This is the $\mathcal{N}=2$ supersymmetric approach's main drawback, indeed, namely that it is limited to capturing interactions
only up to quadratic order in the spins of compact objects,
due to the Grassmann-odd nature of $\psi^{a}$.
While an extension to higher orders in spin is possible in a flat background,
using a set of supercharges $Q_\alpha=p\cdot\psi_\alpha$ with real Grassmann-odd vectors $\psi^a_\alpha(\tau)$
carrying a flavor index $\alpha=1,\ldots,\cN$,
the corresponding $\cN$-fold supersymmetry algebra cannot be realized in an arbitrary curved background spacetime
\cite{Howe:1988ft,Howe:1989vn,Bastianelli:2002qw,Bastianelli:2005vk,Bastianelli:2005uy}.
Fortunately for us, doing so is not necessary in order to describe higher powers of classical spin:
we require only that the covariant SSC constraint $Z^\mu=0$ be preserved,
rather than the more restrictive $\cN$-fold supersymmetry algebra.
The approach of this paper using bosonic oscillators $\alpha^a$ ---
to be introduced momentarily ---
overcomes the inherent limitations of the $\cN=2$ supersymmetric formalism.

\subsection{Introducing the bosonic oscillator approach }\label{sec:oscillators}

Let us now proceed to the main object of this paper:
the introduction of a novel scheme designed to handle spin interactions at any order.
We simply ``bosonize''
the construction of refs.~\cite{Jakobsen:2021lvp,Jakobsen:2021zvh} and introduce a conjugate pair of \emph{commuting}
complex vectors $\alpha^{a}$ and $\bar\alpha^{a}$ living in the tangent space that obey the fundamental 
Poisson brackets\footnote{We note that from a canonical viewpoint a rescaling $\alpha^a\to\alpha^{a}/\sqrt{m}$ would be more natural, as it
provides a canonical Poisson bracket. Yet, we have chosen this convention in order to unify the propagators in the subsequent perturbative treatment.}
\be
 \{\alpha^{a},\bar\alpha^{b}\}
= -\frac{\i}{m} \,\eta^{ab}\, .  \label{funpbosc}
\ee
The spin tensor is defined by 
\be\label{eq:STBO}
S^{\mu\nu}=-2m\i\, e^{\mu}_{a}e^{\nu}_{b}
\bar\alpha^{[a}\alpha^{b]}\, ,
\ee
where we anti-symmetrize with a factor $1/2$. It is a straightforward exercise to check that this
induces the Poisson bracket structure of d'Ambrosi et.~al.~\cite{dAmbrosi:2015ndl,vanHolten:2015vfa,dAmbrosi:2015wqz} given in \eqn{vHposbracks} using $\pi_{\mu}= p_{\mu} - \i \,\omega_{\mu,ab}\bar\alpha^{[a}\alpha^{b]}$
 of \eqn{PiSdef} and the fundamental Poisson brackets $\{x^{\mu}, p_{\nu}\} = \delta^{\mu}_{\nu}$.

A naive counting of degrees of freedom shows that while $S^\mn$ has six, $\alpha^\mu$ and $\balpha^\mu$ together have eight degrees of freedom.
These additional degrees of freedom must be eliminated by internal symmetries of the bosonic oscillators.
Indeed, we find that the physical spin tensor $S^\mn$ is invariant under transformations,
\begin{align}
    \delta_\epsilon \alpha^\mu = \epsilon \balpha^\mu
    \,,
    \qquad
    \delta_{\eps} \balpha^\mu = \bar\epsilon \alpha^\mu 
    \,,
\end{align}
with complex global parameter $\epsilon$.
Due to the complex nature of $\epsilon$, this symmetry exactly removes the two undesired extra degrees of freedom.
We note that the commutator of two such transformations,
\begin{align}
    \delta_{\epsilon}
    \delta_{\epsilon'}
    -        
    \delta_{\epsilon'}
    \delta_{\epsilon}
    =
    \delta_{\lambda}
    \,,
\end{align}
with real parameter $\lambda=2\text{Im}(\epsilon\epsilon')$ is a $U(1)$ rotation:
\begin{align}
    \delta_{\lambda} \alpha^\mu
    =
    \i\lambda \alpha^\mu
    \,,
    \qquad 
    \delta_{\lambda} \balpha^\mu
    =
    -\i\lambda \balpha^\mu
    \,.
\end{align}
The Noether charges of these symmetries are $\alpha^2$, $\balpha^2$ and $\alpha\cdot\balpha$ which ensures conservation of spin length:
\begin{align}
    -\frac12 S^\mn S_\mn 
    =
    (\alpha\cdot\balpha)^2-\alpha^2\balpha^2
    \,,
\end{align}
in perfect agreement with the discussion in section \ref{sec:Hamiltonian}.

The inspiration for this oscillator construction of the spin comes from the bosonic string, which contains an infinite set of higher-spin massive particles 
generated by the worldsheet oscillators $\alpha^{\mu}_{n}(\tau,\sigma)$ upon quantization.  
Our spinning worldline model (in flat space) arises simply as the tensionless
limit ($\alpha'\to \infty$)~\cite{Bengtsson:1986ys,Bouatta:2004kk} of the bosonic
string.  This is achieved upon rescaling the Virasoro generators $L_{0}\to L_{0}/\alpha'$ and $L_{\pm 1}\to L_{\pm 1}/\sqrt{\alpha'}$ that obey an $SL(2,\mathbf{R})$ algebra: in the tensionless limit 
this leaves only $\{x^{\mu},p_{\mu},\alpha_{\pm 1}^{\mu}\}$ in the phase space,
and all string states become massless. This is our phase space.
The transition to massive higher-spin states is performed via a Kaluza-Klein reduction, similar to
the supersymmetric construction~\cite{Bastianelli:2005uy,Jakobsen:2021zvh}.\footnote{We thank Roberto Bonezzi for
crucial discussions on this topic.}
The problem of BRST quantization and equivalence of this bosonic higher-spin particle to higher-spin field
theory was explored in refs.~\cite{Hallowell:2007qk,Bastianelli:2009eh,Bonezzi:2024emt}.

The key advantage of using the bosonic oscillators to model the spin tensor $S^{\mu\nu}$ is that we may now immediately
write down a simple higher-spin worldline Lagrangian that is tailor-made for high-precision perturbative studies at any spin-order.
Following the discussion in the previous
sections, the full SSC-preserving Hamiltonian~\eqref{eq:AllSpinHamiltonian}
expressed in the phase-space variables $\{x^{\mu}, \pi_{\mu}, \alpha^{a},\bar\alpha^{a}\}$ takes the form (to arbitrary orders in curvature)
\be
\label{finalHam}
e^{-1}H_{T}= \sfrac{1}{2} g^{\mu\nu}(x)\, \pi_{\mu}\, \pi_{\nu} - \sfrac{1}{2}m^{2} + \Hnm(x^{\mu}, \pi_{\mu}, \alpha^{a},\bar\alpha^{a})\, ,
\ee
with the einbein $e$ ensuring the reparametrization invariance $\tau\to\tau'(\tau)$.
Note that $\Hnm(x^{\mu}, \pi_{\mu}, \alpha^{a},\bar\alpha^{a})$ is at least linear in curvature, and we express $S^{\mu}$ and
$Z^{\mu}$ in the oscillators using \eqn{aZdefs} and \eqn{eq:STBO}.
Taking into account the fundamental Poisson brackets \eqref{funpxpc} and \eqref{funpbosc},
the first-order form of the action therefore reads
\be
\label{firstorderS}
\tilde S = - \int\!\d\tau\,\Bigl [ p_{\mu}\dot x^{\mu} - m\i \bar\alpha^{a} \dot \alpha^{b}\eta_{ab} -  H_{T}
\Bigr ] \, .
\ee
The algebraic equations of motion
for $p_{\mu}$, i.e.~$\dot{x}^\mu\approx\{x^\mu,H_T\}$, are
\be\label{pcandef}
e^{-1}\dot x^{\mu} = \left[ \pi^{\mu} + \frac{\delta \Hnm(x,\pi,\alpha,\bar\alpha)}{\delta \pi_{\mu}}  \right] _{\pi_{\mu}
=p_{\mu} - \i m \,\omega_{\mu,ab}\bar\alpha^{[a}\alpha^{b]}} \, ,
\ee
using $ \frac{\delta H_{T}}{\delta p_{\mu}} = \frac{\delta H_{T}}{\delta \pi_{\mu}} $. This is an implicit equation allowing us to express $p_{\mu}$ as a function of of $\{x^{\mu},\dot x^{\mu}, \alpha^{a},\bar\alpha^{a}\}$.
It is convenient to choose the proper time gauge and set  $e=1/m$.
Reinserting the (implicit) result for $p_{\mu}(x^{\mu},\dot x^{\mu}, \alpha^{a},\bar\alpha^{a})$ into the first-order Lagrangian \eqref{firstorderS}
yields the worldline Lagrangian expressed in
the variables $\{x,\dot x, \alpha, \dot \alpha, \bar\alpha\}$
\be
\label{sWQFTaction}
S= -\int\!\d\tau\, \Bigl [
\sfrac{m}{2} g_{\mu\nu}\dot x^{\mu}\dot x^{\nu} - m\i \eta_{ab}\bar \alpha^{a} \frac{\D\alpha^{b}}{\d\tau}
-\sfrac{1}{m} \Bigl \{\Hnm +\sfrac{1}{2} \left(\frac{\delta \Hnm}{\delta \pi_{\mu}}\right )^{2} \Bigr \}
  \Bigr ]_{\pi=\pi(x,\dot x, \alpha,\bar\alpha)}
\ee
with the covariant derivative
\be
\frac{\D\alpha^{a}}{\d\tau} = \dot \alpha^{a} - \omega_{\mu}{}^{a}{}_{b}\dot x^{\mu} \alpha^{b} \, .
\ee
Note that we have dropped a non-dynamical term $\frac{m}{2}$. Importantly, in the above the terms 
containing $\Hnm$ and its variations
need to be evaluated for $\pi_{\mu}=\pi_{\mu}(x,\dot x, \alpha,\bar\alpha)$ as the solution of \eqn{pcandef}.
Given this precondition the worldline Lagrangian \eqref{sWQFTaction} is exact.

\subsection{Linear-curvature couplings}

To \emph{linear} order in curvature this solution or \eqn{pcandef} reads
\be \label{psoltoR}
\pi_{\mu}= m\dot x_{\mu} - \frac{\delta \Hnm}{\delta \pi_{\mu}}\Bigr|_{\pi\to m\dot x} + \cO(R^{2})\, ,
\ee
as $\Hnm\sim \cO(R)$:
\be
\Hnm = -B_{SZ}+\sum_{n>1}\tilde{H}_{RS^n} 
+\cO(R^{2})\, .
\ee
Therefore, if we are only interested in the linear-in-curvature terms we may drop
the last term $\left(\frac{\delta \Hnm}{\delta \pi_{\mu}}\right )^{2} $ in the Lagrangian of \eqn{sWQFTaction} and insert $\pi=m\dot x$ for $\Hnm$.
With the linear-in-curvature form of the non-minimal 
Hamiltonian of \eqn{eq:AllSpinHamiltonian}, we therefore find the higher-spin worldline theory action
\begin{align}\label{eq:sWQFTH}
S:=&
 -\int\!\d\tau\, \Bigl [ 
\sfrac{m}{2} g_{\mu\nu}\dot x^{\mu}\dot x^{\nu} - \i m \bar \alpha_{a} \frac{\D\alpha^{a}}{\d\tau}
-m \B_{\ga\gz} \\ & 
-m\sum_{n>1} (1-\i|\dot x|^{-1}\nabla_\gz)\frac{(\nabla_\ga)^{n-2}}{n!}\left.\begin{cases}
		{C}_{ES^n} \E_{\ga\ga}\,,& n\text{ even},\\
		-\i{C}_{BS^n}\B_{\ga\ga}\,,& n\text{ odd},
	\end{cases}
 \right\}
 + \cO(R^{2})
  \Bigr ]\, .\nn
\end{align}
Here we have introduced
\begin{align}\label{aZdefswithosc}
\begin{aligned}
    \E_{\mu\nu}&:=E_{\mu\nu}\big|_{\pi^{\mu}\rightarrow m\dot{x}^{\mu}}\,, &
    \B_{\mu\nu}&:=B_{\mu\nu}\big|_{\pi^{\mu}\rightarrow m\dot{x}^{\mu}}\,, \\
        \ga^{\mu}&:=\frac{\i}mS^{\mu}\big|_{\pi^{\mu}\rightarrow m\dot{x}^{\mu}}=
        \frac{\varepsilon^{\mu\dot{x}\bar{\alpha}\alpha}}{|\dot x|}\,, &
        \gz^{\mu}&:=\frac{\i}mZ^{\mu}\big|_{\pi^{\mu}\rightarrow m\dot{x}^{\mu}}=
        \frac{2\bar{\alpha}^{[\mu}\alpha^{\nu]}\dot{x}_{\nu}}{|\dot x|}\, .
\end{aligned}
\end{align}
The factor of $\i/m$ in the definitions of $\ga^{\mu}$ and $\gz^{\mu}$ cleans up the action by absorbing the signs and factors of the mass appearing in 
e.g.~\cref{eq:AllSpinHamiltonian}.
Up to these phases and factors of the mass, these objects can be thought of as the configuration-space spin and SSC vectors, which differ from the phase-space spin quantities used up to now by curvature corrections.

From \cref{aZdefswithosc} one can see that the decomposition in terms of oscillators imposes an additional restriction on the spin degrees of freedom, namely that $\ga\cdot\gz=0$ --- a condition which need not be obeyed by the spin and SSC vectors as defined in \cref{eq:spinDecomp}.
Evaluating the Poisson brackets of $S\cdot Z$ using \cref{vHposbracks}, one easily finds that $\{H,S\cdot Z\}\sim\cO(Z)$ for any Hamiltonian $H$.
As such, we can be assured that this quantity does not affect dynamics as long as the initial SSC vector is vanishing, justifying the decomposition \cref{eq:STBO} for describing systems with the covariant SSC.

\subsection{Leading higher-curvature couplings}

Based on the Hamiltonian analysis in \cref{sec:SSCR2Ind} we learned that the first induced quadratic-in-curvature terms arise at
the quartic spin order. Let us now translate these additional non-minimal terms to the Lagrangian formulation involving the
bosonic oscillator formalism.

We must first determine how the Hamiltonian terms of various curvature orders enter the Lagrangian in \cref{sWQFTaction}.
Towards this end, we decompose
$\Hnm$ into linear- and quadratic-in-curvature terms: 
\be
\Hnm = \Hnm^{(1)} + \Hnm^{(2)}\, .
\ee
Plugging this into \cref{sWQFTaction} and expanding to quadratic order in curvature, the terms to be added to the Lagrangian \eqn{eq:sWQFTH} take the form 
\be\label{eq:SR2}
S|_{R^{2}} =  -\int\!\d\tau \Bigl [ -\frac{1}{m} 
\Hnm^{(2)}\Bigr|_{\pi\to m\dot x} + \frac{1}{2m}\Big(\frac{\delta \Hnm^{(1)}}{\delta \pi_{\mu}}\Bigr|_{\pi\to m\dot x}\Big)^{2}
\Bigr ]\, .
\ee
We stress that the sign in the last term is not a typo, as inserting the linear in curvature
corrected solution for $\pi_{\mu}= m\dot x_{\mu}+\pi^{(1)}_{\mu}$ with $\pi^{(1)}_{\mu}= - \frac{\delta \Hnm^{(1)}}{\delta \pi_{\mu}}\Bigr|_{\pi\to m\dot x} $ of \eqn{psoltoR}
into $\Hnm^{(1)} $ yields
\begin{align}
\Hnm^{(1)}\Bigr|_{\pi\to m\dot x+ \pi^{(1)}_{\mu}} \!= &
\Hnm^{(1)}\Bigr|_{\pi\to m\dot x}\!\!\! + \pi^{(1)}_{\mu} \frac{\delta \Hnm^{(1)}}{\delta \pi_{\mu}}\Bigr|_{\pi\to m\dot x} %\nn\\
\!\!\!\!\!\!= \Hnm^{(1)}\Bigr|_{\pi\to m\dot x} \!\!\!- \left(\frac{\delta \Hnm^{(1)}}{\delta \pi_{\mu}}\Bigr|_{\pi\to m\dot x}\right )^{2}
\end{align}
giving the total contribution of \eqn{eq:SR2}.
As we have restricted our Hamiltonian analysis to quartic order in the spin tensor in section \ref{sec:SSCR2Gen}, we limit ourselves to the same precision here.
Concretely, then, $\Hnm^{(2)}$ comprises the induced operators in \cref{eq:InducedR2} as well as the operators in \cref{eq:R2S4Ops}.
For the square of the variation of $\Hnm^{(1)}$ within our present quartic-in-spin scope 
we only have the term contributing
\be
    %\Hnm^{(1)}\Bigr |_{S^{2}}= \sfrac{1}{8} R_{\mu\nu\rho\sigma}S^{\mu\nu}S^{\rho\sigma}-
    %\sfrac{C_{ES^{2}}-1}{2} E_{SS} \\
    \Hnm^{(1)}\Bigr |_{S^{2}}=-B_{SZ}-\frac{C_{ES^2}}{2}E_{SS}.
\ee
%using \eqn{RSSidentity} and dropping the $\cO(Z^{2})$ term. 
Varying this quantity, we need %only need the result
\begin{align}
    \frac{\partial B_{SZ}}{\partial\pi_{\mu}}
    &=\left({C^{*}_{S\hat{\pi}Z}}^{\mu}+{C^{*}_{Z\hat{\pi}S}}^{\mu}-2B_{SZ}\hat{\pi}^{\mu}+B_{Z\nu}\epsilon^{\mu\nu\hat{\pi}Z}-B_{S\nu}\epsilon^{\mu\nu\hat{\pi}S}\right)|\pi|^{-1}, \\
    \frac{\delta E_{SS}}{\delta \pi^{\mu}} &= 2 \left (  
          B_{S\nu} \varepsilon^{\mu \nu  \hat{\pi} S}
          + E_{S\nu} \varepsilon^{\mu \nu \hat{\pi} Z} \right) 
          |\pi|^{-1},
\end{align}
from which one deduces 
\begin{align}
\begin{split}\label{eq:H1Squared}
    -\frac{1}{2m}\left(\frac{\partial H^{(1)}_{\rm nm}}{\partial\pi_{\mu}}\Bigr|_{\pi\to m\dot x}\right)^{2}
    &=-m\frac{(C_{ES^{2}}-1)}{\dot{x}^{2}}\left\{(C_{ES^{2}}-1)\E_{\ga\ga}\B_{\ga\gz}\frac{}{}\right. \\
    &\left.+\left[\frac{C_{ES^{2}}-1}{2}\B_{\ga\mu}\B_{\ga\nu}-\E_{\gz\mu}\B_{\ga\nu}\right](\ga^{\mu}\ga^{\nu}-\ga^{2}g^{\mu\nu})\right\},
\end{split}
\end{align}
dropping $\cO(Z^{2})$ terms and applying $\ga\cdot\gz=0$ per \cref{aZdefswithosc}.
Combining this with the quadratic-in-curvature terms of \eqn{eq:R2S4Ops},
\begin{align}\label{eq:H2}
    \frac{1}{m}\Hnm^{(2)}\Bigr |_{\pi\rightarrow m\dot{x}}=-m\frac{(C_{ES^2}-1)}{|\dot x|^2}\E_{\ga\ga}\B_{\ga\gz}+\frac{1}{m}\sum_{n>1}\tilde{H}_{R^2S^n}\Bigr |_{\pi\rightarrow m\dot{x}},
\end{align}
we find the following leading higher curvature contributions to the worldline Lagrangian up to quartic order in 
spin:
\begin{align}
\begin{split}\label{eq:R2S4FromHamiltonian}
    S_{S^{\leq4}R^{2}} &= \int\!\d\tau\left\{-m\frac{C_{ES^{2}}-1}{2\dot{x}^{2}}\left[(C_{ES^{2}}-1)\B_{\ga\nu}\B_{\ga\rho}(\ga^{\nu}\ga^{\rho}-\ga^{2}g^{\nu\rho})\right.\right. \\
    &\left.+2C_{ES^{2}}\E_{\ga\ga}\B_{\ga\gz}-2\E_{\gz\mu}\B_{\ga\nu}(\ga^{\mu}\ga^{\nu}-\ga^{2}g^{\mu\nu})\right] \\
    &\left.+ \frac{1}{m} H_{S^{4}R^{2}}\Bigr |_{\pi\rightarrow m\dot{x}}  + \frac{1}{m} (1 - \i|\dot x|^{-1}\nabla_{\gz}) ( H_{R^{2}} + H_{S^{2}R^{2}})\Bigr |_{\pi\rightarrow m\dot{x}} \right].
\end{split}
\end{align}
Again, the gravito-electromagnetic tensors $\E_{\mu\nu}$ and $\B_{\mu\nu}$ are constructed with
$\pi^{\mu}=m\dot{x}^{\mu}$~\eqref{aZdefswithosc}.

\section{The perturbative WQFT framework }\label{sec:FeynmanRules}

The transition from a Hamiltonian to a Lagrangian, performed in the previous section,
has set the stage for the implementation of a perturbative, Feynman-diagrammatic method for solving the classical equations of motion: the WQFT approach~\cite{Mogull:2020sak,Jakobsen:2021zvh,Jakobsen:2022psy}.
In this section we lay out this framework  and describe the Feynman rules of the theory and their extraction.
Moreover, we discuss the connection to the ansatz-based construction of refs.~\cite{Bern:2020buy,Bern:2022kto,Bern:2023ity,Alaverdian:2024spu}.

As in other formulations of WQFT \cite{Mogull:2020sak,Jakobsen:2021smu,Ben-Shahar:2023djm}, the essential element enabling perturbative calculations is a background field expansion about the solution to the non-interacting system.
When gravitational interactions are present, we expand the metric around flat space as $g_{\mu\nu}=\eta_{\mu\nu}+\kappa h_{\mu\nu}$, with $\kappa=\sqrt{32\pi G}$.
Then, the worldline fields acquire the perturbations
\begin{align}\label{eq:WQFTBGExpansion}
x_{i}^{\mu}(\tau)= b_{i}^{\mu}+ v_{i}^{\mu}\tau + z_{i}^{\mu}(\tau)\, , \qquad
\alpha_{i}^{\mu}(\tau)= \alpha_{-\infty,i}^{ \mu}  + \alpha_{i}^{\prime\, \mu}(\tau)\, ,
\end{align}
where the $\kappa\rightarrow0$ dynamics are described by $z_{i}^{\mu}(\tau)=\alpha_{i}^{\prime\,\mu}(\tau)=0$ and the two worldlines are labelled with $i=1,2$.
The worldline background parameters are $b_i^\mu$ and $v_i^\mu$ and  from those we define the relative impact parameter, $b^\mu$ and Lorentz factor $\gamma$:
\be
b^{\mu}= b_{2}^{\mu}-b_{1}^{\mu}\, , \qquad \gamma=v_{1}\cdot v_{2}\,.
\ee
For each dynamical field, we have a corresponding background variable defined by the field's value at past infinity.
We use a $-\infty$ subscript to denote these background values, and a prime to denote the remaining fluctuation.
Examples include
\begin{subequations}
\begin{align}
    S^\mu_i(\tau)
    &=
    m_i a^\mu_{-\infty,i}
    +
    S^{\prime\mu}_i(\tau)
    \\
    Z_i^\mu(\tau) 
    &= Z_{-\infty,i}^\mu
    +
    Z_i^{\prime\mu}(\tau)
    \,,
    \\
    S_{i}^{\mu\nu}(\tau)
    &=
    S_{-\infty,i}^{\mu\nu}+S_{i}^{\prime\mu\nu}(\tau)
    \,,
\end{align}
\end{subequations}
where, in the first line, we introduced the initial-state Pauli-Lubanski vector $a^\mu_{-\infty, i}=S^\mu_{-\infty,i}/m_i$.
Naturally, background parameters obey similar relations as the dynamical ones, and we have
\begin{align}
    S_{-\infty,i}^{\mn}=-2\i m_{i}\bar\alpha_{-\infty,i}^{[\mu}\alpha_{-\infty,i}^{\nu]}
    \,,
\end{align}
where we note, also, that the distinction between local and covariant indices disappears at past infinity.

\subsection{The Lagrangian: an ansatz-based approach}\label{sec:LagrAnsatz}

The WQFT Feynman rules  arise from the substitution of \cref{eq:WQFTBGExpansion} into an action.
For all-spin covariant-SSC-conserving scattering to linear-in-curvature precision, we have developed a suitable action in \cref{eq:sWQFTH}.
We take this opportunity, though, to present an alternative derivation of that action, adopting instead an ansatz-based approach for the enumeration of $\gz$-dependent operators.
This method was first applied in a similar vein in refs.~\cite{Bern:2020buy,Bern:2022kto}, where actions suited to classical computations using quantum fields were constructed.
These were recently connected to worldline actions in the supplemental analyses of refs.~\cite{Bern:2023ity,Alaverdian:2024spu}.

In contrast to passing through a constrained Hamiltonian analysis, as we have done above, the ansatz for $\gz$-dependent operators starts with a priori unconstrained coefficients.
An appropriate ansatz for the action in this regard is\footnote{\label{fn:ParticleLabels}Worldline labels will be omitted for the time being to avoid notational clutter.}
\begin{align}\label{eq:sWQFT2}
    S&=-m\int{\rm d}\tau\,\left(\frac{1}{2}g_{\mu\nu}\dot{x}^{\mu}\dot{x}^{\nu}-\i\eta_{ab}\bar{\alpha}^{a}\frac{\D\alpha^{b}}{\d\tau}-\cL_{\rm nm}\right),
\end{align}
where $\cL_{\rm nm}=\cL_{\rm nm}^{(a)}+\cL_{\rm nm}^{(Z)}$ and
\begin{subequations}\label{eq:SSCViolLagr}
\begin{align}
    \cL_{\rm nm}^{(a)}
    &=\sum_{n=1}^{\infty}\left(\nabla_{\ga}\right)^{2n-2}\left(\frac{C_{ES^{2n}}}{(2n)!}\E_{\ga\ga}-\frac{\i C_{BS^{2n+1}}}{(2n+1)!}\nabla_{\ga}\B_{\ga\ga}\right), \\
    \cL_{\rm nm}^{(Z)}
    &=\frac{C_{2}^{(Z)}}{2}B_{\ga\gz}-\frac{\nabla_{\gz}}{|\dot{x}|}\sum_{n=2}^{\infty}\left(\nabla_{\ga}\right)^{2n-4}\left(\frac{C^{(Z)}_{2n}}{(2n)!}\nabla_{\ga}\B_{\ga\ga}+\frac{\i C^{(Z)}_{2n-1}}{(2n-1)!}\E_{\ga\ga}\right).
\end{align}
\end{subequations}
The constituents of this action were defined in \eqn{aZdefswithosc}.
\Cref{eq:SSCViolLagr} comprises a basis for linear-in-curvature operators at any spin order with up to one power of $\gz^{\mu}$.\footnote{It turns out that operators with higher powers of $\gz^{\mu}$ are redundant with higher-curvature, lower-$\gz$ operators in a covariant-SSC-preserving theory. This follows from the relation $Z^{\mu}=-im\gz^{\mu}+im\frac{C_{ES^{2}}-1}{\dot{x}^{2}}B_{\ga\nu}\left(\ga^{2}g^{\mu\nu}-\ga^{\mu}\ga^{\nu}\right)+\cO(ZR,R^{2})$ in conjunction with the irrelevance of higher-$Z$ operators in the SSC-preserving Hamiltonian.
The upshot is that $\cO(\gz^{0,1})$ operators are sufficient for describing covariant-SSC-preserving dynamics to all orders in perturbation theory.}
Indeed, alternative operators involving $\gz^{\mu}$ contracted with the curvature components can be related to those labelled by $C_{n}^{(Z)}$ via the differential Bianchi identity -- up to operators which are total time derivatives -- meaning they are redundant.
With this, our ansatz possesses the same number of free Wilson coefficients as the linear-in-curvature quantum action in refs.~\cite{Bern:2020buy,Bern:2022kto}.
The upshot of this section will be the determination of the coefficients $C_{n}^{(Z)}$ through perturbative considerations.
In particular, we will see that the formal-1PM-order conservation of the covariant SSC determines all these coefficients such that \cref{eq:SSCViolLagr} reduces to \cref{eq:sWQFTH}.

As we have done in \cref{sec:SSCR2Ind,sec:SSCR2Gen}, we may extend our scope to higher-curvature operators, which are trivially accommodated in our present approach.
Again, there are two categories of $\cO(R^{2})$ operators we must consider: those with and without a factor of $\gz^{\mu}$.
In the latter category, rather than using the general set in \cref{eq:R2S4Ops}, we content ourselves with restricting to coefficient values reproducing known Kerr observables at one-loop order (see e.g. \cite{Chen:2021kxt,Bautista:2022wjf,Saketh:2023bul,Bautista:2023szu,Aoude:2023dui}).
As mentioned below \cref{eq:R2S4Ops}, we find that the linear-in-curvature action~\eqref{eq:sWQFTH} is sufficient towards this end and therefore does not need any $\gz$-independent quadratic-in-curvature operators.
Interestingly, this means that our action~\eqref{eq:sWQFTH} manifests the vanishing of static linear Love numbers for Kerr at quartic order in spin \cite{Ivanov:2022qqt}.

In the $\gz$-dependent-operator category, we include all possibilities at quartic order in the spin tensor with unfixed coefficients:
\begin{align}\label{eq:R2Za3}
\begin{split}
    S_{R^{2}Za^{3}}=-m\int{\rm d}\tau\,\frac{1}{\dot{x}^{2}}&\left[\E_{\ga\mu}\B_{\gz\nu}\left(
    C_{R^{2},1}^{(Z)}\ga^{2}\,g^{\mu\nu}+C_{R^{2},2}^{(Z)}\ga^{\mu}\ga^{\nu}\right)\right. \\
    &\left.+\E_{\gz\mu}\B_{\ga\nu}\left(C_{R^{2},3}^{(Z)}\ga^{2}\,g^{\mu\nu}+C_{R^{2},4}^{(Z)}\ga^{\mu}\ga^{\nu}\right)\right].
\end{split}
\end{align}
Recall from \cref{aZdefswithosc} that $\ga\cdot\gz=0$, restricting the operators that can be written. 
Analogously to $C_{n}^{(Z)}$ at linear order in curvature, we seek to recover \cref{eq:R2S4FromHamiltonian} by determining constraints on the $C_{R^{2},n}^{(Z)}$ from conservation of the covariant SSC, this time at the formal-2PM order.
Most generally, one should include operators with $\cO(\ga^{n<3}\gz)$ in \cref{eq:R2Za3} and determine their coefficients as well.
However, we take advantage of our foreknowledge from \cref{eq:R2S4FromHamiltonian} that no such operators are needed to conserve the covariant SSC to ignore them from the outset; constraining the coefficients in \cref{eq:R2Za3} will be enough of a proof of principle.
An in-depth exploration of quadratic-in-curvature operators at higher spin orders and for general compact objects is left to future work.

Having collected all the ingredients with which we will perform perturbative computations ---
\cref{eq:sWQFT2,eq:SSCViolLagr,eq:R2Za3} --- we move now to the Feynman rules of the theory,
focusing on those pertaining to the worldline and bosonic-oscillator perturbations rather than the well-known bulk gravitational rules.

\subsection{Propagators}

Propagators are derived from the kinetic part of the action defined from the $\kappa\to0$ limit of the action after insertion of the background field expansion~\eqref{eq:sWQFT2}:
\begin{align}
	\left.S\right|_{\rm kinetic}&=
  -m\int\!\d\tau\,\eta_{\mu\nu}\!\left(\frac12\dot z^\mu\dot z^{\nu}
	-\i\bar{\alpha}^{\prime\mu}\dot\alpha^{\prime\nu}\right)\,.
\end{align}
It is convenient to work in momentum (energy) space and we define
\begin{align}\label{eq:1dfourier}
    z^\mu(\tau)=\int_\omega e^{-\i\omega\tau}z^\mu(\omega)\,, &&
    {\alpha^{\prime}}^\mu(\tau)=
    \int_\omega e^{-\i\omega\tau}{\alpha^{\prime}}^\mu(\omega)\,,
\end{align}
with integration measures absorbing factors of $2\pi$: $\int_\omega=\int\sfrac{\d\omega}{2\pi}$.

We employ the in-in formalism discussed in the context of WQFT in~\cite{Jakobsen:2023oow}.
As shown there, in the classical limit, the only difference to the in-out formalism is the use of retarded propagators instead of the (time-symmetric) Feynman prescription.
Propagators thus have a causality associated with them and are not symmetric under $\omega\to-\omega$, as they would be with a Feynman prescription.

For notational convenience, we collect the worldline fields in a composite vector,
\begin{align}
 \Z^{\mu}_{I}(\omega)=\{z^{\mu}(\omega), \alpha^{\prime \mu}(\omega),
\bar\alpha^{\prime \mu}(\omega)\}
\, ,
\end{align}
with flavor index $I$.
The retarded propagator of this flavoured worldline field then reads 
\be
 \begin{tikzpicture}[baseline={(current bounding box.center)}]
    \coordinate (in) at (-0.6,0);
    \coordinate (out) at (1.4,0);
    \coordinate (x) at (-.2,0);
    \coordinate (y) at (1.0,0);
    \draw [zParticle] (x) -- (y) node [midway, below] {$\omega$};
    \draw [dotted] (in) -- (x);
    \draw [dotted] (y) -- (out);
    \draw [fill] (x) circle (.08) node [above] {$\mu$} node [below] {$I$};
    \draw [fill] (y) circle (.08) node [above] {$\nu$} node [below] {$J$};
  \end{tikzpicture}=%\langle \Z_{I}^{\mu}(-\omega) \Z^{\nu}_{J}(\omega) \rangle_{0}=
-\i\frac{\eta^{\mu\nu}}{m}
 \left (\begin{matrix}
  \frac1{(\omega+\i0)^{2}} & & \\
   & &   -\frac1{\omega+\i0} \\
    & \frac1{\omega+\i0} & 
 \end{matrix}\right )_{IJ} \, ,
    \ee
with all other entries vanishing. 
In this diagram, the arrow indicates causality flow and the energy $\omega$ is assumed to be labelled in the same direction as this flow.
Instead, if the labelling would be reversed, essentially sending $\omega\to-\omega$, the $i\eps$ of the propagators, essentiallly, change sign:
\be
 \begin{tikzpicture}[baseline={(current bounding box.center)}]
    \coordinate (in) at (-0.6,0);
    \coordinate (out) at (1.4,0);
    \coordinate (x) at (-.2,0);
    \coordinate (y) at (1.0,0);
    \draw [zParticle] (x) -- (y) node [midway, below] {$-\omega$};
    \draw [dotted] (in) -- (x);
    \draw [dotted] (y) -- (out);
    \draw [fill] (x) circle (.08) node [above] {$\mu$} node [below] {$I$};
    \draw [fill] (y) circle (.08) node [above] {$\nu$} node [below] {$J$};
  \end{tikzpicture}=
-\i\frac{\eta^{\mu\nu}}{m}
 \left (\begin{matrix}
  \frac1{(\omega-\i0)^{2}} & & \\
   & &   \frac1{\omega-\i0} \\
    & -\frac1{\omega-\i0} & 
 \end{matrix}\right )_{IJ} \, .
\ee
Note also that, in contrast to the propagators for Grassmann-odd oscillators, the propagators for the bosonic oscillators are sensitive to the alignment of oscillator flow from $\alpha_{i}^{\mu}$ to $\bar\alpha_{i}^{\mu}$.

\subsection{Interaction Vertices}

The next ingredients for perturbative calculations are the interaction vertices.
While any desired vertex can be derived directly from the action, here we lay out an elegant approach generating vertices with multi-worldline perturbation legs through the action of Poisson brackets on pure-graviton vertices --- echoing our use of brackets in the Hamiltonian analysis.

In order to derive the vertices, we expand the interaction part of the action in the perturbative fields.
The interaction part reads,
\begin{align}
    S_{\rm int}
    &=
    -m\int \d\tau
    \Big(
        h_\mn 
        \dot x^\mu 
        \dot x^\nu 
        +
        \omega^{bc}_\mu S_{bc} \dot x^\mu
    \Big)
    +
    S_{\rm nm}
    =    \int \d \tau
    \cL_{\rm int}
    \,,\nn
\end{align}
where we also introduced its Lagrange function $\cL_{\rm int}$.

Let us first expand the action in the gravitational perturbations, which we write schematically as
\begin{align}\label{eq:ActionExpandedGraviton}
    \cL_{\rm int}
    =
    \sum_{n=1}^\infty
    \int_{k_1\dots k_n} 
    V^{\mnone\dots\mnn}_{(n)}(
        \dot x^\mu,
        \alpha^\mu,
        \balpha^\mu
        ;k_1^\mu\dots k_n^\mu)
    e^{\i x(\tau)\cdot\sum_n k_n}
    h_\mnone(k_1)
    \dots
    h_\mnn(k_n )
    \ .
\end{align}
Here, we work in momentum space for the gravitons.
This step requires one to expand all the gravitational fields used in the action such as the metric, curvature and its covariant derivatives, which is cumbersome but straightforward using computer algebra.
Importantly all dependence on the trajectory $x^\mu(\tau)$ is in the exponential function which simply follows from the fact that the trajectory appears only as the argument of the gravitational fields in position space.
The expansion coefficients $V_{(n)}^{\mnone \dots \mnn}$, then, depend only on the $n$ graviton momenta and polynomially on the velocity $\dot x(\tau)$ and bosonic oscillators $\alpha^\mu(\tau)$ and $\balpha^\mu(\tau)$ (in fact, only through the combination $S^\mn=-2\i m\balpha^{[\mu}\alpha^{\nu]}$).
Clearly, a further expansion in the perturbations of $x^\mu$, $\alpha^\mu$ and $\balpha^\mu$ is now trivial requiring only an expansion of the exponential function and the polynomials in the expansion coefficients.
It is also clear that no information is lost upon inserting the background parameters into the expansion coefficients (as long as the background parameters are kept generic).
Thus, the functions $V_{(n)}^{\mnone\dots\mnn}(v^\mu,\alpha_{-\infty}^{\mu},\balpha_{-\infty}^{\mu};k_1^\mu\dots k_n^\mu)$ contain all required information about $n$-point graviton vertices with any number of outgoing worldline fields.

Feynman rules with outgoing worldline perturbations are now derived by further variations of eq.~\eqref{eq:ActionExpandedGraviton} with respect to the said fields.
Such variations are, however, easily rephrased as partial derivatives with respect to the background parameters.
Thus, the operators for attaching an external $z$, $\alpha'$ or $\bar\alpha'$ leg are (see also refs.~\cite{Jakobsen:2023oow,Kim:2024svw})
\begin{align}\label{eq:GenericDiagrams}
    &V_{n\mid I_{1}\dots I_{m}}^{
        \mu_1\nu_1\dots\mu_n\nu_n|
        \sigma_1\dots\sigma_m
        } =
    \resizebox{0.25\textwidth}{!}{
\begin{tikzpicture}[baseline={(current bounding box.center)}]
    \coordinate (in) at (-1,0);
    \coordinate (out1) at (1.2,0);
    \coordinate (out2) at (1.2,0.5);
    \coordinate (out3) at (1.2,0.9);
    \coordinate (x) at (0,0);
      \node (k1) at (-1.2,-1.5) {$\mu_{1}\nu_{1}$};
    \node (k2) at (1.2,-1.5) {$\mu_n\nu_{n}$};
    \node (w2) at (0.2,1,0) {$\omega_{m}$};
    \draw (out1) node [right] {$\Z_{I_1}^{\sigma_1}$};
    \draw (out2) node [right] {$\!\!\!\vdots$};
     \draw (-.2,-0.8) node [right] {$\!\!\!\ldots$};
    \draw (out3) node [right] {$\Z_{I_m}^{\sigma_m}$};
    \draw [worldlineStatic] (in) -- (x) ;
    \draw [zParticle] (x) -- (out1)  node [midway, above] {$\omega_{1}$};
    \draw [zParticle] (x) to[out=70,in=180] (out3) ;
    \draw [graviton] (x) -- (k1) node [midway, left] {$k_{1}\,$};
      \draw [graviton] (x) -- (k2) node [midway, right] {$\,k_{n}$};
    \draw [fill] (x) circle (.08);
    \end{tikzpicture}
    }
      \\&\qquad
      =
      \dd\left(\sum_{i=1}^{m} \omega_i +\sum_{j=1}^{n}v\cdot k_i\right)
      \left (\prod_{i=1}^m \cD_{I_i}^{\sigma_i}\right )
      e^{\i b\cdot \sum_{j}^{n} k_j}
      V^{\mnone\dots\mnn}_{(n)}[
          v,\alpha_{-\infty},\bar\alpha_{-\infty}
          ;
          k_1\dots k_n
      ]
      \nn
\end{align}
where the operators $\cD_I^\sigma$ can be expressed in terms of Poisson brackets taken with respect to the background variables:
\begin{align}
    \cD_1^\sigma V 
    =
    m\{-v^\sigma+\i\omega b^\sigma,V\}
    \,,\qquad 
    \cD_2^\sigma V
    =
    \i m\{\alpha_{-\infty}^\sigma,V\}
    \,,\qquad 
    \cD_3^\sigma V
    =
    -\i m\{\balpha^\sigma_{-\infty},V\}
    \,.
\end{align}
The background Poisson brackets follow from the $\tau\to-\infty$ limit of the generic brackets, Eqs.~\eqref{vHposbracks} and~\eqref{funpbosc}, and read
\begin{subequations}\label{eq:BackgroundPoissonBrackets}
\begin{align}
    \{ 
        b^\mu ,
        v^\nu 
        \} 
        &=\frac{1}{m}
        \eta^{\mu\nu}
        \,,
        \\
    \{ 
        \alpha^\mu_{-\infty} ,
        \balpha^\nu_{-\infty}
        \} 
        &=
        -\frac{\i}{m} 
        \eta^{\mu\nu}
        \,,
\end{align}
\end{subequations}
with all other brackets zero.

In order to use this recursion the parameters in $V^{\mnone\dots\mnn}_{(n)}[
          v,\alpha_{-\infty},\bar\alpha_{-\infty}
          ;
          k_1\dots k_n
      ]$ must be left unconstrained, e.g.~$v^{2}=1$ and $v\cdot k=0$ -- where $k^{\mu}:=\sum_{i=1}^{n}k_{i}^{\mu}$ -- should not be imposed.
     For example, the one-graviton generator is:
\begin{align}
\begin{aligned}\label{eq:OneGravGenerator}
    &V_{(1)}^{\mu\nu}[v,\alpha,\bar\alpha;k]
    =-\i m\left\{\frac{\kappa}{2}v^{\mu}v^{\nu}
    +\frac{\i\kappa}{2m}(k\cdot S)^{(\mu}v^{\nu)}
    +\frac{1}{v^{4}}\left[\frac{ Z^{\beta}}{m}\frac{\partial\tilde{C}^{*(1)}_{av Z v}(-k)}{\partial h_{\mu\nu}(-k)}\right.\right. 
    \\
    &
    +\sum_{n=1}^{\infty}\left(\frac{k\cdot a}{\sqrt{v^{2}}}\right)^{2n-2}\left(C_{ES^{2n}}
    +\frac{C^{(Z)}_{2n
    +1}}{2n
    +1}\frac{\i k\cdot Z}{m\sqrt{v^{2}}}\right)\frac{1}{(2n)!}\frac{\partial\tilde{C}^{(1)}_{avav}(-k)}{\partial h_{\mu\nu}(-k)} 
    \\
    &\left.\left.
        +\sum_{n=1}^{\infty}\left(\frac{k\cdot a}{\sqrt{v^{2}}}\right)^{2n-1}\left(C_{BS^{2n
    +1}}
    +\frac{C^{(Z)}_{2n
    +2}}{2n
    +2}\frac{\i k\cdot Z}{m\sqrt{v^{2}}}\right)\frac{\i}{(2n
    +1)!}\frac{\partial\tilde{C}^{*(1)}_{avav}(-k)}{\partial h_{\mu\nu}(-k)}\right]\right\}.
\end{aligned}
\end{align}
To compactify the expression, we have omitted $-\infty$ labels on the background variables $\alpha^\mu_{-\infty}$ $\bar\alpha^\mu_{-\infty}$, $S^\mn_{-\infty}$, $a^\mu_{-\infty}$, and $Z^\mu_{-\infty}$.
Additionally, we have written the graviton dynamics implicitly as the partial derivative of the momentum-space Weyl tensor expanded to linear order in $\kappa$, $\tilde{C}^{(1)}_{\alpha\beta\rho\tau}(-k)$:
\begin{align}
\begin{aligned}\label{eq:DerivWeylTensor}
    \frac{\partial\tilde{C}^{(1)}_{\alpha\beta\rho\tau}(-k)}{\partial h_{\mu\nu}(-k)}&=-\kappa k_{[\alpha}\eta_{\beta][\rho}\delta_{\tau]}^{(\mu}k^{\nu)}-\kappa k_{[\rho}\eta_{\tau][\alpha}\delta_{\beta]}^{(\mu}k^{\nu)}-\kappa\delta_{[\alpha}^{(\mu}(k_{\beta]}k_{[\rho}-k^{2}\eta_{\beta][\rho})\delta_{\tau]}^{\nu)} \\
    &+\frac{\kappa}{3}\eta^{\mu\nu}(k^{2}\eta_{\alpha[\rho}\eta_{\tau]\beta}+3k_{[\rho}\eta_{\tau][\alpha}k_{\beta]}+\eta_{\alpha[\tau}\eta_{\rho]\beta}k^{\mu}k^{\nu}).
\end{aligned}
\end{align}
Substituting this into \cref{eq:GenericDiagrams}, we can now generate any Feynman rule with one graviton and any number of worldline perturbations.
For example, the one-graviton Feynman rules with no worldline perturbations is (for brevity, we print the following rules for $Z_{-\infty}^{\mu}=0$)
\begin{align}
    &V_{1|}^{\mu\nu}=\vcenter{\hbox{\begin{tikzpicture}[line cap=round,line join=round,x=1cm,y=1cm]
        \node (A1) at (-3.1,-.8) {\footnotesize{$\mu\nu$}};
        \path [draw=black, worldlineStatic] (-4.2,0) -- (-2.7,0);
        \path [draw=black, graviton] (-3.45,0) -- (-3.45,-1.);
         \draw [fill] (-3.45,0) circle (.08);
    \end{tikzpicture}}} \\
    &=-\i m\dd(v\cdot k)e^{\i b\cdot k}\left[\frac{\kappa}{2}v^{\mu}v^{\nu}+\frac{\i\kappa}{2m}(k\cdot S)^{(\mu}v^{\nu)}\right.\nn \\
    &\left.+\sum_{n=1}^{\infty}\left(k\cdot a\right)^{2n-2}\left(\frac{C_{ES^{2n}}}{(2n)!}\frac{\partial\tilde{C}^{(1)}_{avav}(-k)}{\partial h_{\mu\nu}(-k)}+(k\cdot a)\frac{\i C_{BS^{2n+1}}}{(2n+1)!}\frac{\partial\tilde{C}^{*(1)}_{avav}(-k)}{\partial h_{\mu\nu}(-k)}\right)\right].\nn
\end{align}
Again, we have omitted explicit $-\infty$ subscripts in this equation.
More complicated rules involving one graviton and one worldline perturbation are included in \Cref{app:1WLFR}.

Let us end this section with a word of caution concerning dimensional regularization and
the presence of Levi-Civita-tensors --- being defined in $D=4$ --- 
in the worldline vertices. As we are only allowing parity-even interactions the total 
number of Levi-Civitas in any vertex is even. There are two schemes in dealing with this issue
in dimensional regularization where we lift our theory to $D=4-2\epsilon$ dimensions: (a) 
Contract all Levi-Civita products to $\eta_{\mu\nu}$'s and upgrade these thereafter to $D$ dimensions.
(b) Contract all Levi-Civita products and keep the resulting $\eta_{\mu_{4}\nu_{4}}$'s in $D=4$ dimensions, thereby
splitting the indices as $\mu \to (\mu_{4}, \mu_{\epsilon})$. It is crucial to do so
consistently in a calculation. The final (renormalized) results will not depend on this
choice, yet intermediate results can be scheme dependent.

\subsection{Observables from the WQFT formalism}

The Lagrangian ansatz in \cref{eq:SSCViolLagr} can now be readily applied to the perturbative computation of observables.
After quickly reviewing the role of one-point functions in WQFT, we discuss the constraints on the Lagrangian ansatz emerging from the requirement that the covariant SSC is perturbatively conserved up to one-loop order.
The upshot is that we will recover the precise action derived via the Hamiltonian analysis in \Cref{sec:Hamiltonian,sec:TransitionToLagrangian}.

As in the case of WQFT constructed using Grassmann variables, observables are related to one-point functions of the dynamical perturbations in the WQFT action \cite{Mogull:2020sak,Jakobsen:2021smu,Jakobsen:2021lvp,Jakobsen:2021zvh}.
Specifically, reinstating particle labels, the linear impulse and spin kick depend on the worldline and oscillator perturbations respectively through\footnote{We do not differentiate between locally-flat and curved spacetime indices here as observables are measured at flat time-like infinity.}\
\begin{subequations}\label{eq:ImpulseSpinKick}
\begin{align}
    \Delta p^{\mu}_{i}&=-m_i\omega^{2}\left.\langle z^{\mu}_{i}(-\omega)\rangle\right|_{\omega\rightarrow0}\,, \\
    \Delta \alpha^{\mu}_{i}&=\i\,\omega\left.\langle\alpha^{\prime\,\mu}_{i}(-\omega)\rangle\right|_{\omega\rightarrow0}\,.
\end{align}
\end{subequations}
Here $\Delta p_{i}^{\mu}$ is the change of momentum, or impulse, of black hole $i$ under
the scattering, and $ \Delta \alpha^{\mu}_{i}$ is the change in the oscillator worldline
field. The latter induces the change in the spin tensor, or spin kick, $ \Delta S_i^{\mu\nu}$ under
the scattering via
\begin{align}\label{eq:SpinKick}
    \Delta S_i^{\mu\nu}=-2\i m_i(\bar{\alpha}^{[\mu}_{-\infty,i}\Delta\alpha^{\nu]}_i+\Delta\bar{\alpha}^{[\mu}_i\alpha^{\nu]}_{-\infty,i}+\Delta\bar{\alpha}^{[\mu}_i\Delta\alpha^{\nu]}_i) \, .
\end{align}
In a two-body context, the one-point function $\langle \Z_{I,i}^{\mu}(-\omega)\rangle$ is computed perturbatively via the diagram \cite{Mogull:2020sak}
\begin{align}\label{eq:1PtFns}
    \langle \Z_{I,i}^\mu(-\omega)\rangle=\vcenter{\hbox{\begin{tikzpicture}[line cap=round,line join=round,x=1cm,y=1cm]
        \node (A1) at (-1.33,.0) {$\Z_{I,i}^\mu(-\omega)$};
        \node (A2) at (-2.5,-.3) {$\omega\rightarrow$};
        \path [draw=black, worldlineStatic] (-4.7,0) -- (-3.83,0);
        \path [draw=black, zParticle] (-3.17,0) -- (-2.2,0);
        \path [draw=black, worldlineStatic] (-4.7,-1.5) -- (-3.83,-1.5);
        \path [draw=black, worldlineStatic] (-3.17,-1.5) -- (-2.2,-1.5);
        %\path [draw=black, zUndirected] (-4.6,0) -- (-3.4,0);
        \draw[pattern=north east lines] (-3.5,-0.75) ellipse (.55cm and .9cm);
    \end{tikzpicture}}},
\end{align}
where the striped ellipse represents a sum over all connected diagrams where the two worldlines fluctuate and interact gravitationally. As we are interested in the classical ($\hbar \to 0$) result, only
tree-level graphs are to be considered.
The overall factors of $\omega$ in \cref{eq:ImpulseSpinKick} cancel the propagator for the external leg and thus amount to an amputation of the outgoing leg.
When observables are extracted directly through these one-point functions --- rather than by passing through the eikonal, for example \cite{Jakobsen:2021zvh} --- we are allowed to impose the initial conditions on the parameters at the level of the Feynman rules, i.e. that the initial SSC vector is vanishing.

\subsection{Perturbative SSC preservation}

Having reviewed the extraction of observables from WQFT, we turn our focus to the change in the SSC vector.
As a composition of the spin tensor and the worldline velocity, its change can be constructed from \cref{eq:ImpulseSpinKick}, \eqref{eq:SpinKick} most generally through
\begin{align}
    \Delta Z_{i}^{\mu}&=\frac{1}{m}\Delta p_{i,\nu}S_{-\infty,i}^{\mu\nu}+v_{i,\nu}\Delta S_{i}^{\mu\nu}+\frac{1}{m}\Delta p_{i,\nu}\Delta S_{i}^{\mu\nu}.\label{eq:DeltaZ}
\end{align}
However, as our interest lies in the conservation of the covariant SSC, we can replace $\Delta Z_{i}^{\mu}=0$ with a simpler yet equivalent condition.
The key insight here is that the change in the bosonic oscillator and its conjugate can be unified through
\begin{align}\label{eq:ChangeInOsc}
    \Delta\alpha^{\mu}_{i}&=\alpha^{\rho}_{-\infty,i}{f_{i,\rho}}^{\mu}\,,&
    \Delta\bar\alpha^{\mu}&=\bar\alpha^{\rho}_{-\infty,i}{f_{i,\rho}}^{\mu},
\end{align}
for some common function ${f_{i,\rho}}^{\mu}$ whose precise form will not concern us.
This is a consequence of the fact that $\alpha^{\mu}_{-\infty,i}$ and $\balpha^{\mu}_{-\infty,i}$ enter observables only through the spin tensor.
The change in spin tensor then takes the alternative form
\begin{align}
    \Delta S^{\mu\nu}_{i}&=-2S_{-\infty,i}^{\rho[\mu}{f_{i,\rho}}^{\nu]}+S_{-\infty,i}^{\tau\rho}{f_{i,\tau}}^{\mu}{f_{i,\rho}}^{\nu}\,,
\end{align}
expressed directly in terms of the initial spin tensor.
Plugging this in to \cref{eq:DeltaZ}, and requiring conservation of the covariant SSC perturbatively,
we learn that
\begin{align}\label{eq:EquivalenceSSCConservation}
\begin{split}
    \Delta Z_{i}^{\mu}=0\quad &
    \Leftrightarrow\quad \Delta(\dot{x}_{i}\cdot\alpha_{i})=0\quad \Leftrightarrow\quad \Delta(\dot{x}_{i}\cdot\bar\alpha_{i})=0.
\end{split}
\end{align}
Thus, conservation of the SSC vector is equivalent to conservation of the 
``bosonic supersymmetry'' charges  $p_{i}\cdot \alpha_{i}$ and
 $p_{i}\cdot \balpha_{i}$ in our theory.

Moving now to the perturbative computation, at leading order we must evaluate
\begin{align}\label{eq:LO1PtFns}
    \langle \Z_{I,i}^\mu(-\omega)\rangle_{\rm LO}=\vcenter{\hbox{\begin{tikzpicture}[line cap=round,line join=round,x=1cm,y=1cm]
        \node (A1) at (-1.33,.0) {$\Z_{I,i}^\mu(-\omega)$};
        \node (A2) at (-2.5,-.3) {$\omega\rightarrow$};
        \path [draw=black, worldlineStatic] (-4.7,0) -- (-3.45,0);
        \path [draw=black, zParticle] (-3.45,0) -- (-2.2,0);
        \path [draw=black, worldlineStatic] (-4.7,-1.5) -- (-2.2,-1.5);
        \path [draw=black, photon] (-3.45,0) -- (-3.45,-1.5);
         \draw [fill] (-3.45,0) circle (.08);
    \draw [fill] (-3.45,-1.5) circle (.08);
    \end{tikzpicture}}},
\end{align}
for $I=1,2,3$.
Imposing $\left.\Delta Z_{i}^{\mu}\right|_{\rm LO}=0$, all coefficients of $\gz^{\mu}$-dependent operators in \cref{eq:SSCViolLagr} are constrained to obey 
\begin{align}\label{eq:1PMSSCConservation}
    C_{2}^{(Z)}&=2\,,&
    C_{n}^{(Z)}&=n\times\begin{cases}
    C_{ES^{n-1}}, & n\text{ odd}, \\
    C_{BS^{n-1}}, & n\text{ even},
    \end{cases}
\end{align}
where $n\geq 3$.
Substituting these back into \cref{eq:SSCViolLagr}, our ansatz reduces to \cref{eq:sWQFTH} which was derived from the Hamiltonian analysis.
Additionally, we have confirmed that when the Kerr values $C_{i,(E/B)S^{n}}=1$ are chosen, we find agreement with the impulse and spin kick derived from other methods \cite{Vines:2017hyw,Guevara:2018wpp,Guevara:2019fsj,Aoude:2021oqj}.

In order to constrain the quadratic-in-curvature operators~\eqref{eq:R2Za3},
we move to next-to-leading order.
Besides the leading-order one-point functions~\eqref{eq:LO1PtFns},
which contribute to the change in SSC vector through the last term in \cref{eq:DeltaZ},
we need the sub-leading one-point functions:
\begin{align}\label{eq:NLO1PtFns}
    &\langle \Z_{I,i}^\mu(-\omega)\rangle_{\rm NLO} \\
    &\qquad=%\vcenter{\hbox{\begin{tikzpicture}[line cap=round,line join=round,x=1cm,y=1cm]
    \begin{tikzpicture}[baseline={([yshift=-.5ex](.5,-.75))}]
        %%%%%%%%%%%%%%%%%%%%%%
        %\node (A1) at (2.37,.0) {$\Z_{I,i}^\mu(-\omega)$};
        %\node (A2) at (1.2,-.3) {$\omega\rightarrow$};
        \path [draw=black, worldlineStatic] (-1.5,0) -- (-0.6,0);
        \path [draw=black, zParticle] (0.6,0) -- (1.5,0);
        \path [draw=black, worldlineStatic] (-1.5,-1.5) -- (1.5,-1.5);
        \path [draw=black, zParticle] (-.6,0) -- (.6,0);
        \path [draw=black, photon] (-0.6,-1.5) -- (-0.6,0);
        \path [draw=black, photon] (0.6,-1.5) -- (0.6,0);
        \filldraw[fill=black] (-0.6,0) circle (3pt);
        \filldraw[fill=black] (-0.6,-1.5) circle (3pt);
        \filldraw[fill=black] (0.6,0) circle (3pt);
        \filldraw[fill=black] (0.6,-1.5) circle (3pt);
    \end{tikzpicture}
    +
    \frac12
    \begin{tikzpicture}[baseline={([yshift=-.5ex](.5,-3.75))}]
        %%%%%%%%%%%%%%%%%%%%%%
        \path [draw=black, worldlineStatic] (-3.5,-3) -- (-.5,-3);
        \path [draw=black, zParticle] (-2,-3) -- (-.5,-3);
        \path [draw=black, worldlineStatic] (-3.5,-4.5) -- (-.5,-4.5);
        \path [draw=black, photon] (-2.6,-4.5) -- (-2,-3);
        \path [draw=black, photon] (-1.4,-4.5) -- (-2,-3);
        \filldraw[fill=black] (-2,-3) circle (3pt);
        \filldraw[fill=black] (-2.6,-4.5) circle (3pt);
        \filldraw[fill=black] (-1.4,-4.5) circle (3pt);
    \end{tikzpicture}
    +
    \frac12
    \begin{tikzpicture}[baseline={([yshift=-.5ex](.5,-3.75))}]
        %%%%%%%%%%%%%%%%%%%%%%
        \path [draw=black, worldlineStatic] (.5,-3) -- (3.5,-3);
        \path [draw=black, zParticle] (2,-3) -- (3.5,-3);
        \path [draw=black, worldlineStatic] (.5,-4.5) -- (3.5,-4.5);
        \path [draw=black, photon] (2,-3.75) -- (2,-3);
        \path [draw=black, photon] (1.4,-4.5) -- (2,-3.75);
        \path [draw=black, photon] (2.6,-4.5) -- (2,-3.75);
        \filldraw[fill=black] (2,-3) circle (3pt);
        \filldraw[fill=black] (2,-3.75) circle (3pt);
        \filldraw[fill=black] (1.4,-4.5) circle (3pt);
        \filldraw[fill=black] (2.6,-4.5) circle (3pt);
    \end{tikzpicture}
    \nn
    \\
    &\qquad+\begin{tikzpicture}[baseline={([yshift=-.5ex](.5,-.75))}]
        %%%%%%%%%%%%%%%%%%%%%%
        \path [draw=black, worldlineStatic] (-1.5,-1.5) -- (-0.6,-1.5);
        \path [draw=black, zParticle] (-0.6,-1.5) -- (0.6,-1.5);
        \path [draw=black, worldlineStatic] (0.6,-1.5) -- (1.5,-1.5);
        \path [draw=black, worldlineStatic] (-1.5,0) -- (0.6,0);
        \path [draw=black, zParticle] (.6,0) -- (1.5,0);
        \path [draw=black, photon] (-0.6,-1.5) -- (-0.6,0);
        \path [draw=black, photon] (0.6,-1.5) -- (0.6,0);
        \filldraw[fill=black] (-0.6,0) circle (3pt);
        \filldraw[fill=black] (-0.6,-1.5) circle (3pt);
        \filldraw[fill=black] (0.6,0) circle (3pt);
        \filldraw[fill=black] (0.6,-1.5) circle (3pt);
    \end{tikzpicture}
    +
    \begin{tikzpicture}[baseline={([yshift=-.5ex](.5,-.75))}]
        %%%%%%%%%%%%%%%%%%%%%%
        \path [draw=black, worldlineStatic] (-1.5,-1.5) -- (1.5,-1.5);
        \path [draw=black, worldlineStatic] (-1.5,0) -- (0.6,0);
        \path [draw=black, zParticle] (.6,0) -- (1.5,0);
        \path [draw=black, photon] (0,-1.5) -- (-0.6,0);
        \path [draw=black, photon] (0,-1.5) -- (0.6,0);
        \filldraw[fill=black] (-0.6,0) circle (3pt);
        \filldraw[fill=black] (0.6,0) circle (3pt);
        \filldraw[fill=black] (0,-1.5) circle (3pt);
    \end{tikzpicture}
    +
    \begin{tikzpicture}[baseline={([yshift=-.5ex](.5,-.75))}]
        %%%%%%%%%%%%%%%%%%%%%%
        \path [draw=black, worldlineStatic] (-1.5,-1.5) -- (1.5,-1.5);
        \path [draw=black, worldlineStatic] (-1.5,0) -- (0.6,0);
        \path [draw=black, zParticle] (.6,0) -- (1.5,0);
        \path [draw=black, photon] (0,-0.75) -- (-0.6,0);
        \path [draw=black, photon] (0,-0.75) -- (0.6,0);
        \path [draw=black, photon] (0,-0.75) -- (0,-1.5);
        \filldraw[fill=black] (-0.6,0) circle (3pt);
        \filldraw[fill=black] (0.6,0) circle (3pt);
        \filldraw[fill=black] (0,-0.75) circle (3pt);
        \filldraw[fill=black] (0,-1.5) circle (3pt);
    \end{tikzpicture}\,.
    \nn
\end{align}
Internal $\cZ^{\nu}_{J,j}$ propagators here involve a sum over the flavor index $J$.
As these calculations involving one-loop integrals are now standard,
we refer the reader to refs.~\cite{Mogull:2020sak,Jakobsen:2021zvh} for further computational details.
Assembling any of the three conditions in \cref{eq:EquivalenceSSCConservation} at next-to-leading order
determines the Wilson coefficients in \cref{eq:R2Za3}:
\begin{align}
    &C_{R^{2},1}^{(Z)}=0,\quad C_{R^{2},2}^{(Z)}=C_{ES^{2}}(C_{ES^{2}}-1),\quad C_{R^{2},3}^{(Z)}=-C_{R^{2},4}^{(Z)}=C_{ES^{2}}-1.%,\quad\text{\kaystext{BSZ operators}} \\
\end{align}
Thus, we reproduce precisely the quadratic-in-curvature part of the action as derived from the covariant-SSC-preserving Hamiltonian in \cref{eq:R2S4FromHamiltonian}.
Along the way, we have verified that we reproduce the known NLO linear impulse in the Kerr limit up to fourth order in the spin vector \cite{Bautista:2023szu,Aoude:2023dui}.
In summary: we have demonstrated the compatibility of the Hamiltonian analysis in \Cref{sec:Hamiltonian,sec:TransitionToLagrangian} with perturbative computations of observables from an ansatz-based Lagrangian.

\section{Conclusions}

The WQFT formalism has proven itself a remarkably powerful tool for calculating
scattering observables in classical GR.
Significant calculations of scattering observables ---
the scattering angle, fluxes, spin kick, the gravitational waveform~\cite{Jakobsen:2021smu,Jakobsen:2021lvp} ---
have been done involving spinning massive bodies,
up to quadratic order in spins at 3PM~\cite{Jakobsen:2022fcj,Jakobsen:2022zsx},
and linear in spins at 4PM~\cite{Jakobsen:2023ndj,Jakobsen:2023hig}.
These results are already being used in effective-one-body (EOB) models,
both to describe large-angle scattering~\cite{Rettegno:2023ghr,Buonanno:2024vkx}
and bound-orbit waveforms~\cite{Buonanno:2024byg}.
However, the use of an $\cN=2$ supersymmetric worldline action has so far
meant a restriction to \emph{quadratic} powers in spin (quadrupoles) ---
a restriction that other approaches involving scattering amplitudes have not had~\cite{Guevara:2018wpp,Chung:2018kqs,Guevara:2019fsj,Bern:2020buy,Chen:2021kxt,Aoude:2022trd,Bern:2022kto,Bautista:2023szu,Aoude:2023dui,Chen:2024mmm,Bohnenblust:2024hkw}.
This restriction follows from the need to ensure that all spin degrees of freedom
are captured by a three-dimensional Pauli-Lubanski spin vector $\vec{S}$,
imposing a spin-supplementary condition (SSC) on the relativistic spin tensor $S^{\mu\nu}$.
The $\cN=2$ formulation enforces this SSC using a corresponding pair of conserved supercharges.
Unfortunately, as there exists no covariant gravitational coupling
of the supersymmetric worldline action beyond $\cN=2$ supersymmetries,
further progress has been hindered ---
although, an alternative WQFT-based approach using a spinning body-fixed frame
has recently been introduced~\cite{Ben-Shahar:2023djm}.

In this paper, we laid the foundations for an efficient higher-spin WQFT implementation,
designed for future calculations at high orders in perturbation theory.
The essential challenge was to build an SSC-preserving worldline action,
suitable for WQFT-style quantization,
that avoids the inherent limitations of the $\cN=2$ SUSY action.
Our starting point was instead the Hamiltonian,
which depends only on the particle's essential degrees of freedom ---
position $x^\mu$, spin $S^{\mu\nu}$, and the metric $g_{\mu\nu}$ ---
and wherein the SSC manifests itself as a second-class constraint on the dynamics
(in the language of Dirac~\cite{dirac2001lectures}).
By explicitly solving for the SSC,
a suitable ansatz could be made for the Hamiltonian depending only on the spin vector $S^\mu$
(instead of the spin tensor $S^{\mu\nu}$)
both with linear and quadratic curvature couplings.
In the corresponding first-order worldline action,
related to the Hamiltonian by a simple Legendre transform,
we then showed how there arises the need to enlarge the spinning phase space.
Motivated by the desire for a covariant SSC constraint $\pi_\mu S^{\mu\nu}=0$,
we found that a convenient encoding
is via a set of complex bosonic oscillators $\alpha^a$, with $S^{ab}=-2\i m\bar{\alpha}^{[a}\alpha^{b]}$.
Inspiration for this encoding comes from the bosonic string,
our model arising in the tensionless limit.
The constructed bosonic worldline action may be used inside or outside the WQFT framework.

Alongside our work on the Hamiltonian, we demonstrated an alternative
 ansatz-based approach for obtaining the SSC-conserving worldline action.
As its starting point, this method took an ansatz for the most general action constructed from the worldline degrees of freedom with up to one power of $\gz^{\mu}$.
The number of free Wilson coefficients in this action corresponds to that of the linear-in-curvature action of refs.~\cite{Bern:2020buy,Bern:2022kto}.
To connect to the Hamiltonian analysis, we evaluated the perturbative change in the SSC vector and determined conditions on the Wilson coefficients of the action such that this change vanishes.
In this way, coefficients were fixed for all linear-in-curvature operators and for the leading (in terms of the physical PM counting) quadratic in curvature operators.
As a by-product of this analysis, we validated that this novel action produces known Kerr observables at leading and subleading loop orders \cite{Vines:2017hyw,Guevara:2019fsj,Bautista:2023szu,Aoude:2023dui}.

Having now developed a higher-spin WQFT framework,
there exist several opportunities for follow-up calculations.
Most natural would be to extend the scattering observables at formal 3PM order (two loops)
to cubic and quartic order in spin, which would have a physical 6PM and 7PM counting respectively. 
Seeing as the associated Hamiltonian for the conservative dynamics would be fully local in time
(i.e.~avoid any nonlocalities arising from tails),
such results would have immediate relevance for bound two-body motion.
At 4PM order a quadratic-in-spin result (physical 6PM) would also be new:
given our present discussion of higher-curvature terms we anticipate a $1/\eps$ dim-reg divergence,
to be renormalized by adding a suitable counter term to the action~\cite{Jakobsen:2023pvx}.
The leading-order gravitational waveform can also be generalized to include higher-spin effects,
a calculation already done using amplitudes-based methods~\cite{DeAngelis:2023lvf,Brandhuber:2023hhl,Aoude:2023dui,Brandhuber:2024qdn}.

In the longer term, the main application for these perturbative scattering results will likely
be within EOB models
(or similar resummations) used to describe both unbound and bound two-body motion in the strong-field regime.
From that perspective, the eikonal phase --- that we have not focused on in this paper ---
will likely play a crucial future role.
In its guise as the radial action, a gauge-invariant quantity,
it is straightforward to interpolate the eikonal phase the two-body and (deformed) one-body systems.
Extending our ability to calculate the eikonal phase to higher PM orders using WQFT methods
will therefore be an important task,
one that will also allow us to leverage powerful unitarity-based methods for future calculations.
We leave this exciting prospect for future work.

\section*{Acknowledgments}
We thank M.~Ben-Shahar, R.~Bonezzi, I.~Costa, J.~W.~Kim, R. Patil, M.V.S. Saketh and J.~Steinhoff for discussions and 
important comments.
G.M.~is supported by the UK Royal Society under grant URF\textbackslash R1\textbackslash 231578,
``Gravitational Waves from Worldline Quantum Field Theory''.
This work was funded by the Deutsche Forschungsgemeinschaft
(DFG, German Research Foundation)
Projektnummer 417533893/GRK2575 ``Rethinking Quantum Field Theory'' 
and  by the European Union through the 
European Research Council under grant ERC Advanced Grant 101097219 (GraWFTy).
Views and opinions expressed are however those of the authors only and do not necessarily reflect those of the European Union or European Research Council Executive Agency. Neither the European Union nor the granting authority can be held responsible for them.
This research was also supported by the Munich Institute for Astro-, Particle and BioPhysics (MIAPbP) which is funded by the Deutsche Forschungsgemeinschaft (DFG, German Research Foundation) under Germany´s Excellence Strategy – EXC-2094 – 390783311.

\appendix

\section{Feynman rules with worldline perturbations}\label{app:1WLFR}

For illustrative purposes, we collect here the Feynman rules with one graviton and one worldline perturbation generated through \cref{eq:GenericDiagrams} from the one-graviton generator \cref{eq:OneGravGenerator}.
The condition $Z_{-\infty}^{\mu}=0$ has again been imposed on the rules in this section to simplify them and all background quantities are taken at infinite past, though we omit the $-\infty$ subscripts on $\alpha_{-\infty}^{\mu}$, $\bar\alpha_{-\infty}^{\mu}$, $a^\mu_{-\infty}$ and $S_{-\infty}^\mn$.
The vertex involving one graviton and one trajectory perturbation represents the rule
\begin{align}
    &V_{1|1}^{\mu\nu|\rho}=
    \vcenter{\hbox{\begin{tikzpicture}[line cap=round,line join=round,x=1cm,y=1cm]
        \node (A1) at (-3.1,-.8) {\footnotesize{$\mu\nu$}};
        \node (A2) at (-1.9,-0.) {\footnotesize{$\cZ_{1}^{\rho}(-\omega)$}};
        \path [draw=black, worldlineStatic] (-4.2,0) -- (-3.45,0);
        \path [draw=black, zParticle] (-3.45,0) -- (-2.7,0);
        \path [draw=black, graviton] (-3.45,0) -- (-3.45,-1.);
         \draw [fill] (-3.45,0) circle (.08);
    \end{tikzpicture}}} \\
    &=m\dd(v\cdot k+\omega)e^{\i b\cdot k}\left\{\frac{\kappa}{2}v^{\mu}v^{\nu}k^{\rho}+\frac{\i\kappa}{2m}(k\cdot S)^{(\mu}v^{\nu)}k^{\rho}+\frac{\omega}{m} S^{\tau\rho}\frac{\partial\tilde{C}^{*(1)}_{av\tau v}(-k)}{\partial h_{\mu\nu}(-k)}\right.\nn \\
    &+\sum_{n=1}^{\infty}\left(k\cdot a\right)^{2n-3}\left[k^{\rho}\left(k\cdot a\right)\left(\frac{C_{ES^{2n}}}{(2n)!}\frac{\partial\tilde{C}^{(1)}_{avav}(-k)}{\partial h_{\mu\nu}(-k)}+(k\cdot a)\frac{\i C_{BS^{2n+1}}}{(2n+1)!}\frac{\partial\tilde{C}^{*(1)}_{avav}(-k)}{\partial h_{\mu\nu}(-k)}\right)\right.\nn \\
    &+\frac{\i\omega}{m}(k\cdot S)^{\rho}\left(k\cdot a\right)\left(\frac{C^{(Z)}_{2n+1}}{(2n+1)!}\frac{\partial\tilde{C}^{(1)}_{avav}(-k)}{\partial h_{\mu\nu}(-k)}+\frac{\i C^{(Z)}_{2n+2}}{(2n+2)!}(k\cdot a)\frac{\partial\tilde{C}^{*(1)}_{avav}(-k)}{\partial h_{\mu\nu}(-k)}\right)\nn \\
    &-\frac{\omega}{m}\epsilon^{\sigma\lambda\alpha\beta}{P_{v,\lambda}}^{\rho} S_{\alpha\beta}\left[\left(k\cdot a\right)\left(\frac{C_{ES^{2n}}}{(2n)!}\frac{\partial\tilde{C}^{(1)}_{\sigma vav}(-k)}{\partial h_{\mu\nu}(-k)}+\frac{\i C_{BS^{2n+1}}}{(2n+1)!}\left(k\cdot a\right)\frac{\partial\tilde{C}^{*(1)}_{\sigma vav}(-k)}{\partial h_{\mu\nu}(-k)}\right)\right.\nn \\
    &\left.-\frac{1}{2}k_{\sigma}\left((2n-2)\frac{C_{ES^{2n}}}{(2n)!}\frac{\partial\tilde{C}^{(1)}_{avav}(-k)}{\partial h_{\mu\nu}(-k)}+(2n-1)\frac{\i C_{BS^{2n+1}}}{(2n+1)!}\left(k\cdot a\right)\frac{\partial\tilde{C}^{*(1)}_{avav}(-k)}{\partial h_{\mu\nu}(-k)}\right)\right]\nn \\
    &\left.\left.+2\omega v^{\alpha}P_{v}^{\beta\rho}\left(k\cdot a\right)\left(\frac{C_{ES^{2n}}}{(2n)!}\frac{\partial\tilde{C}^{(1)}_{a\alpha a\beta}(-k)}{\partial h_{\mu\nu}(-k)}+\frac{\i C_{BS^{2n+1}}}{(2n+1)!}(k\cdot a)\frac{\partial\tilde{C}^{*(1)}_{a\alpha a\beta}(-k)}{\partial h_{\mu\nu}(-k)}\right)\right]\right\}.\nn
\end{align}
Here we have employed the projector $P_{v}^{\mu\nu}=\eta^{\mu\nu}-v^{\mu}v^{\nu}$.
Replacing the trajectory perturbation with an oscillator perturbation, \cref{eq:GenericDiagrams} gives instead
\begin{align}
    &V_{1|2}^{\mu\nu|\rho}=
    \vcenter{\hbox{\begin{tikzpicture}[line cap=round,line join=round,x=1cm,y=1cm]
        \node (A1) at (-3.1,-.8) {\footnotesize{$\mu\nu$}};
        \node (A2) at (-1.9,-0.) {\footnotesize{$\cZ_{2}^{\rho}(-\omega)$}};
        \path [draw=black, worldlineStatic] (-4.2,0) -- (-3.45,0);
        \path [draw=black, zParticle] (-3.45,0) -- (-2.7,0);
        \path [draw=black, graviton] (-3.45,0) -- (-3.45,-1.);
         \draw [fill] (-3.45,0) circle (.08);
    \end{tikzpicture}}} \\
    &=2m\dd(v\cdot k+\omega)e^{\i b\cdot k}\left\{\frac{\i\kappa}{4}v^{(\mu}\left[(k\cdot\alpha)\eta^{\nu)\rho}-k^{\rho}\alpha^{\nu)}\right]-a^{\sigma}\eta^{\rho[\tau}\alpha^{\nu]}v_{\nu}\frac{\partial\tilde{C}^{*(1)}_{\sigma v\tau v}(-k)}{\partial h_{\mu\nu}(-k)}\right.\nn \\
    &-\i k_{\tau}\eta^{\rho[\tau}\alpha^{\nu]}v_{\nu}\sum_{n=1}^{\infty}\left(k\cdot a\right)^{2n-2}\left(\frac{C^{(Z)}_{2n+1}}{(2n+1)!}\frac{\partial\tilde{C}^{(1)}_{avav}(-k)}{\partial h_{\mu\nu}(-k)}+\frac{\i C^{(Z)}_{2n+2}}{(2n+2)!}(k\cdot a)\frac{\partial\tilde{C}^{*(1)}_{avav}(-k)}{\partial h_{\mu\nu}(-k)}\right)\nn \\
    &+\epsilon^{\sigma v\rho\tau}\alpha_{\tau}\sum_{n=1}^{\infty}\left(k\cdot a\right)^{2n-3}\left[\left(k\cdot a\right)\left(\frac{C_{ES^{2n}}}{(2n)!}\frac{\partial\tilde{C}^{(1)}_{\sigma vav}(-k)}{\partial h_{\mu\nu}(-k)}+\frac{\i C_{BS^{2n+1}}}{(2n+1)!}(k\cdot a)\frac{\partial\tilde{C}^{*(1)}_{\sigma vav}(-k)}{\partial h_{\mu\nu}(-k)}\right)\right.\nn \\
    &\left.\left.+\frac{1}{2}k_{\sigma}\left((2n-2)\frac{C_{ES^{2n}}}{(2n)!}\frac{\partial\tilde{C}^{(1)}_{avav}(-k)}{\partial h_{\mu\nu}(-k)}+(2n-1)\frac{\i C_{BS^{2n+1}}}{(2n+1)!}(k\cdot a)\frac{\partial\tilde{C}^{*(1)}_{avav}(-k)}{\partial h_{\mu\nu}(-k)}\right)\right]\right\}.\nn
\end{align}
The rule for a perturbation of the conjugate oscillator is related to this one through
\begin{align}
    V^{\mu\nu|\rho}_{1|3}&=\vcenter{\hbox{\begin{tikzpicture}[line cap=round,line join=round,x=1cm,y=1cm]
        \node (A1) at (-3.1,-.8) {\footnotesize{$\mu\nu$}};
        \node (A2) at (-1.9,-0.) {\footnotesize{$\cZ_{3}^{\rho}(-\omega)$}};
        \path [draw=black, worldlineStatic] (-4.2,0) -- (-3.45,0);
        \path [draw=black, zParticle] (-3.45,0) -- (-2.7,0);
        \path [draw=black, graviton] (-3.45,0) -- (-3.45,-1.);
         \draw [fill] (-3.45,0) circle (.08);
    \end{tikzpicture}}}=V^{\mu\nu|\rho}_{1|2}\Big|_{\alpha\rightarrow-\bar\alpha}.
\end{align}
These involve functional derivatives of the momentum-space Weyl curvature tensor expanded to linear order in $\kappa$, which has been given in \cref{eq:DerivWeylTensor}.

\bibliographystyle{JHEP}
\bibliography{busy.bib}

\end{document}